\documentclass[]{aastex63}

\usepackage{xspace, float, amssymb, amsmath, xcolor, fancyhdr, graphicx,physics, tabularx, CJK}
\newcommand{\AM}{\textsc{AM$^3$}\xspace}

\newcommand{\equ}[1]{Eq.~\ref{equ:#1}}

\received{}
\revised{}
\accepted{}

\shorttitle{AM$^3$: Open Source Lepto-Hadronic Modeling}
\shortauthors{Klinger et al.}

\begin{document}
\begin{CJK*}{UTF8}{gbsn}

\title{AM$^3$: An Open-Source Tool for Time-Dependent Lepto-Hadronic Modeling of Astrophysical Sources}

\correspondingauthor{Marc Klinger}
\email{marc.klinger@desy.de}

\correspondingauthor{Annika Rudolph}
\email{mail@annikarudolph.de}

\correspondingauthor{Xavier Rodrigues}
\email{xavier.rodrigues@eso.org}

\correspondingauthor{Shan Gao}
\email{mathgaoshanphy@gmail.com}

\author[0000-0002-4697-1465]{Marc Klinger}
\affil{Deutsches Elektronen-Synchrotron DESY, 
Platanenallee 6, 15738 Zeuthen, Germany}

\author[0000-0003-2040-788X]{Annika Rudolph}
\affil{Niels Bohr International Academy and DARK, 
Niels Bohr Institute, University of Copenhagen, 
Blegdamsvej 17, \\
2100, Copenhagen, Denmark}

\author[0000-0001-9001-3937]{Xavier Rodrigues}
\affil{European Southern Observatory, Karl-Schwarzschild-Straße 2, 85748 Garching bei München, Germany}

\author[0000-0003-0327-6136]{Chengchao Yuan (袁成超)}
\affil{Deutsches Elektronen-Synchrotron DESY, 
Platanenallee 6, 15738 Zeuthen, Germany}

\author[0000-0003-1143-3883]{Ga\"{e}tan Fichet de Clairfontaine}
\affil{Julius-Maximilians-Universit\"{a}t W\"{u}rzburg, Fakult\"{a}t f\"{u}r Physik und Astronomie, Emil-Fischer-Str. 31, D-97074 W\"{u}rzburg, Germany}

\author[0000-0003-2837-3477]{Anatoli Fedynitch}
\affiliation{Institute of Physics, Academia Sinica, Taipei City, 11529, Taiwan}
\affil{Institute for Cosmic Ray Research, the University of Tokyo, 5-1-5 Kashiwa-no-ha, Kashiwa, Chiba, 277-8582, Japan}

\author[0000-0001-7062-0289]{Walter Winter}
\affil{Deutsches Elektronen-Synchrotron DESY, 
Platanenallee 6, 15738 Zeuthen, Germany}

\author[0000-0001-7861-1707]{Martin Pohl}
\affil{Deutsches Elektronen-Synchrotron DESY, 
Platanenallee 6, 15738 Zeuthen, Germany}
\affil{Institute of Physics and Astronomy, University of Potsdam, 14476 Potsdam, Germany}

\author[0000-0002-5309-2194]{Shan Gao}
\affil{Deutsches Elektronen-Synchrotron DESY, 
Platanenallee 6, 15738 Zeuthen, Germany}

\affil{Sartorius Corporate Administration GmbH, 
Otto-Brenner-Strasse 20, 30379, G\"{o}ttingen, Germany}

\begin{abstract}
We present the \AM (``Astrophysical Multi-Messenger Modeling'') software.
\AM is a documented open source software 
\footnote{\label{footnote:source_code}The BSD-licensed source code can be found at \url{https://gitlab.desy.de/am3/am3} \citep[for v1.1.0 at Zenodo see][]{zenodo_ref}, accompanied by a comprehensive user guide and documentation at \url{https://am3.readthedocs.io/en/latest/}.} 
that efficiently solves the coupled integro-differential equations describing the temporal evolution of the spectral densities of particles interacting in astrophysical environments, including photons, electrons, positrons, protons, neutrons, pions, muons, and neutrinos. The software has been extensively used to simulate the multi-wavelength and neutrino emission from active galactic nuclei (including blazars), gamma-ray bursts, and tidal disruption events.
The simulations include all relevant non-thermal processes, namely synchrotron emission, inverse Compton scattering, photon-photon annihilation, proton-proton and proton-photon pion production, and photo-pair production. The software self-consistently calculates the full cascade of primary and secondary particles, including non-linear feedback processes and predictions in the time domain. It also allows to track separately the particle densities produced by means of each distinct interaction processes, including the different hadronic channels.
With its efficient hybrid solver combining analytical and numerical techniques, \AM combines efficiency and accuracy at a user-adjustable level. 
We describe the technical details of the numerical framework and present three examples of applications to different astrophysical environments. 
\end{abstract}

\section{Introduction}

In the era of multi-messenger astroparticle physics, our understanding of high-energy astrophysical sources builds on a growing wealth of data not only across the electromagnetic spectrum, but also in the cosmic-ray and neutrino domains. Evidence for a high-energy astrophysical neutrino flux was first obtained by the IceCube neutrino observatory in the South Pole \citep{IceCube:2013low}. High-energy neutrinos are known to be produced when cosmic rays can interact (lepto-)hadronically, either in the astrophysical source where they were accelerated, or en route to Earth. Neutrinos are therefore a crucial component in the quest to identify the sites of cosmic-ray acceleration.

In recent years, the multi-messenger community has been looking to shed light on the nature of the high-energy neutrino sources. For this we rely on continued neutrino observations, multi-wavelength monitoring, and follow-up programs. At the same time, we need a phenomenological understanding of the conditions surrounding cosmic-ray interactions, in order to relate neutrino and multi-wavelength emission. To that end, an efficient numerical framework to simulate cosmic-ray interactions is an indispensable tool.

Currently, one of the most promising object classes is that of blazars, which are active galactic nuclei (AGN) whose relativistic jet points toward earth. In recent years, there have been multiple associations between high-energy neutrino events and individual blazars~\citep[e.g.,][]{2016NatPh..12..807K,Gao:2016uld,2019ApJ...880..103G, IceCube:2018dnn,Giommi:2020viy,2020ApJ...902...29P,Franckowiak:2020qrq,Sahakyan:2022nbz,2023ApJ...954...70A,deClairfontaine:2023pgo}. Several tidal disruption events (TDEs) have also been suggested as potential counterpart sources of IceCube neutrinos \citep[e.g.,][]{2021NatAs...5..510S, 2022PhRvL.128v1101R, 2021arXiv211109391V}. More recently, the Seyfert galaxy NGC~1068 has also been identified as a neutrino source candidate at the 4.2~$\sigma$ confidence level by the IceCube collaboration~\citep{IceCube:2022der}.
On the other hand, the community is still awaiting a multi-messenger observation with $\gtrsim 5 \sigma $ confidence.

The above observations have triggered theoretical modelling and analysis efforts ~\citep[e.g.,][]{Cerruti:2018tmc,Gao:2018mnu,Keivani:2018rnh,Sahu:2020eep,Rodrigues:2020fbu,deClairfontaine:2023pgo,Yuan_2023,2024arXiv240611513Y}. More broadly speaking, the nature of the high-energy emission from astrophysical source is also a hotly debated topic. In particular, cosmic-ray proton interactions have long been suggested as potential contributors to the X-ray and $\gamma$-ray fluxes from AGN and gamma-ray bursts \citep[GRBs; see, e.g.,][]{Mannheim:1993jg,Dermer:2016jmw,Madejksi:2016}.

Numerical AGN studies suggest that non-linear electromagnetic cascades inside the source can lead to highly non-trivial multi-wavelength signatures of cosmic-ray interactions~\citep[e.g.,][]{Reimer:2018vvw,Petropoulou:2022sct,Rodrigues:2023vbv}. This means that the relationship between cosmic rays, neutrinos, and multi-wavelength photon emission is a complex one that relies on sophisticated, self-consistent numerical interaction models.

Accurate, physics-driven multi-messenger modeling is also essential in preparing for the next generation of neutrino experiments such as KM3NeT \citep{KM3Net:2016zxf}, IceCube~Gen2 \citep{IceCube-Gen2:2020qha}, Baikal-GVD \citep{Allakhverdyan:2021vkk}, P-ONE \citep{P-ONE:2020ljt}, GRAND~\citep{Kotera:2021hbp}, RNO-G~\citep{RNO-G:2020rmc}, HUNT \citep{Huang:2023mzt}, TAMBO \citep{TAMBO:2023plw}, TRIDENT \citep{TRIDENT:2023} or NEON \citep[e.g.,][]{Zhang:2023icn}, which will increase our neutrino detection sensitivity and significantly push the high-energy energy boundary. New and upcoming multi-wavelength observatories
such as ULTRASAT \citep{Sagiv:2013rma}, ZTF \citep{Graham:2019qsw}, Vera Rubin Observatory \citep{LSST:2008ijt}, SVOM \citep{Wei:2016eox}, CTA \citep{CTAConsortium:2017dvg} and LHAASO \citep{Cao:2010zz}, among others, will further boost the number of neutrino associations, and provide increasingly refined data that require more precise modeling.

When performing time-dependent modeling, we self-consistently follow the evolution of particles interacting in an astrophysical environment. In this approach, the primary particles, which are those originally accelerated, are subject to interactions with the same magnetic and photon fields as the secondary particles, which are the ones produced as a consequence of those interactions.
A range of software follows this method: in the AGN context, well-known examples include \textsc{ATHE$\nu$A} \citep{Mastichiadis:1994hq, Mastichiadis:2005nj, Dimitrakoudis:2012uf, Petropoulou:2014lja}, \AM \citep[][hereafter G16]{Gao:2016uld}, \textsc{SOPRANO} \citep{Gasparyan:2021oad}, \textsc{Katu} \citep[][open source]{Jimenez-Fernandez:2021pdx} and \textsc{LeHaMoc} \citep[][whose leptonic simulation component is currently open-source software]{Stathopoulos:2023qoy}. Further examples have been presented by \citet{1992MNRAS.258..657C,1998ApJS..114..269N,2005ApJ...628..857P,Asano:2007my, Asano:2008tc,2008A&A...491..617B, Vurm:2008ue}.
In contrast, software such as that described in \citet{Boettcher:2013wxa} or \textsc{LeHa-Paris} \citep{Cerruti:2014iwa} uses a steady-state solution for the primaries to compute the full secondary cascade. Recently, the Hadronic Code Comparison Project has demonstrated a good agreement between different numerical approaches \citep{Cerruti:2021hah}.

In past years, the role of open-source software has gained importance in high-energy astroparticle physics. First, because software re-usability improves scientific results and speeds up research due to efficient resource management. Secondly, because open-source code bases enable the reproduction and cross-checking of published results, and make it easier for research groups to learn from each other. Finally, they are an integral part of open science, as specified by the UNESCO Open Science Recommendation from 2017 \footnote{\url{https://www.unesco.org/en/open-science}} and endorsed by the European Commission.

In this paper we present \AM (based on G16), an open-source software to self-consistently compute the temporal evolution of energy spectra of photons, electrons, positrons, protons, neutrons, neutrinos and intermediate species. As mentioned previously, secondary particles undergo the same reactions as primaries. The software includes all relevant (non-thermal) leptonic and hadronic processes, while offering the option to separately track the photon and neutrino signatures of each process separately. 
In addition to an efficient hybrid solver, there are available options for further performance optimization, which can reduce the computational time to $\mathcal{O}$(3~s) for a lepto-hadronic AGN model. The software has been applied to lepto-hadronic modeling of AGN \citep{Rodrigues:2018tku,Rodrigues:2020fbu,MAGIC:2023ybm,Rodrigues:2023vbv,deClairfontaine:2023pgo}, GRB prompt \citep{Rudolph:2021cvn, Rudolph:2022ppp, Rudolph:2022dky, Rudolph:2023auv} as well as afterglow emission \citep{2024arXiv240313902K}, and TDEs \citep{Yuan_2023,2024arXiv240109320Y,2024arXiv240611513Y}.

This paper is structured as follows: in Sec.~\ref{sec:overview} we provide an overview of the code, including the form of the system of differential equations, the particle interactions channels and other physical processes included in the software, the basic structure of the numerical solver, and the basic input/output framework available to the user. In Sec.~\ref{sec:applications} we provide examples of applications to astrophysical environments, in Sec.~\ref{sec:efficiency} we briefly discuss the software performance, and in Sec.~\ref{sec:conclusions} we present our conclusions. 
Further quantitative details on the physical processes and the numerical solver are provided in App.~\ref{app:physics_processes} and~\ref{app:solver}, respectively. There we also provide details on the options available for performance optimizing.

\section{Code overview}
\label{sec:overview}

In this section we introduce the main aspects of \AM, namely the type of equation system being solved, the physical processes included, the numerical solver methods, and a brief description of the user interface. 
Broadly speaking, \AM is designed to calculate non-thermal spectra of high-energy astrophysical sources. To this end, it solves the coupled temporal evolution of protons, neutrons, electrons, positrons, photons, neutrinos, muons, and pions as a function of energy. It does not include a treatment of nuclei heavier than protons.
All particles are assumed to be within a homogeneous and isotropic radiation zone. Particles injected by the user in the system are denoted as ``primaries", whereas ``secondary particles" are those created through interactions throughout the simulation. 

\subsection{General form of the equation system}
\label{sec:equation_system}

The following paragraphs closely follow the description in G16, where the code structure was introduced. Generally, the particle kinematics are described by a set of coupled integro-differential equations, derived from each species' Boltzmann equation, coupled via collision terms. \AM focuses on non-thermal relativistic particle spectra, where the approximation of particle energy $E$ being proportional to momentum $p$ holds, $E \approx pc$, with the speed of light $c$. Rewriting the Boltzmann equation in terms of energy, the partial differential equations (PDEs) take the following conceptual form:

\begin{equation} 
	\partial_{t}n(E,t)=-\partial_{E} \qty[ \dot{{E}}({E,t})n(E,t) ] -\alpha(E,t) n(E,t)+Q(E,t) \, .
	\label{equ:overalllinearform}
\end{equation}
Here, $n(E,t)$ is the differential number density for energy $E$ and time $t$ and the total number of particles equals $N=\int \dd V\int \dd E \, n(E)$.
The right-hand side of the equation contains terms representing continuous cooling/advection [$\dot{{E}}({E,t})>0$], injection/source [$Q(E,t)$] and escape/sink [$\alpha(E,t)$]. 
Sink and source terms may be obtained as follows.

For a process of $a+b\rightarrow c+d$, the generation rate (or source term) of particle species $c$ can be written as an integration over the parent particle population $b$:

\begin{equation}
	Q_{c}(E_{c})=\int \dd E_{a}\: n_{a}(E_{a}) \:\int \dd E_{b}\: n_{b}(E_{b}) \: R_\mathrm{a,b\rightarrow c}(E_a, E_b, E_c)   \, .
	\label{eqn:Qc}
\end{equation}
where $R_{a,b\rightarrow c}(E_a, E_b, E_c)$ is the differential cross-section $\dd\sigma_{a,b\rightarrow c} / \dd E_{c} \dd\mu$ of generating the particle $c$ with $E_{c}$, averaged over the cosine $\mu$ of the reaction angle $\theta$, that can be defined, e.g., between the incident particle $b$ and outgoing particle $c$ in the lab frame: 

\begin{equation}
	R_\mathrm{a,b\rightarrow c}(E_a, E_b, E_c) = \frac{c}{2}\int \dd \mu \: (1-\mu) \: \frac{\dd\sigma_{a,b\rightarrow c}}{\dd E_{c} \dd\mu}(E_{c},E_{b},E_{a},\mu) \, .
\end{equation}

For a process like $b+a\rightarrow b+c$, the effect on the particle $b$ can be treated either through a continuous cooling term (thus, $\dot{E}$) or via discrete sink and source terms (sometimes also called ``discrete energy losses''). In the latter case, the particle $b$ reappears with a different energy and the particle $b$ on the left-hand side is nevertheless treated as ``annihilated'' here. The disappearance rate for $b$, is therefore

\begin{equation}
	\alpha_b(E_{b})=\int \dd E_a \: n_{a}(E_{a}) \: \int \dd E_c \: R_\mathrm{a,b\rightarrow c} (E_a, E_b, E_c) 
\end{equation}
and the re-appeared $b$ after the reaction is treated as a new particle. The corresponding re-appearance rate can be obtained through Eq.~\ref{eqn:Qc}.

\subsection{Overview of interaction processes}

\AM includes the treatment of the following processes:
\begin{itemize}
    \item Synchrotron radiation of all charged particles.
    \item Inverse Compton scatterings between all charged particles and photons.
    \item Photon-photon pair annihilation.
    \item Photo-hadronic (Bethe-Heitler) electron-positron pair production. 
    \item Proton-photon pion production.
    \item Proton-proton pion production.
    \item Adiabatic expansion.
    \item Pion and muon decay.
    \item Escape and injection. 
\end{itemize}

The terms entering Eq.~\ref{equ:overalllinearform} for each of these processes are summarized in Table~\ref{tab:coefficients}. From here on we use the standard symbols $p$ for protons, $n$ for neutrons, $\pi^\pm$ for charged pions, $\pi^0$ for neutral pions, $\mu^\pm_{R/L}$ for right ($R$) or left ($L$) handed muons, $\nu_{e/\mu}$/$\bar{\nu}_{e/\mu}$ for electron ($e$) or muon($\mu$) neutrinos/anti-neutrinos, $e^\mp$ for electrons/positrons and $\gamma$ for photons.
For the details on the implementation of each process, as well as available options for performance optimization, we refer the interested reader to App.~\ref{app:physics_processes}. 

\begin{deluxetable}{|c|c|c|c|c|c|c|c|c| }
    \tablewidth{0pt}
    \tablecaption{List of coefficients.\label{tab:coefficients}}
        \tablehead{ & $ e^{-}$  & $ e^{+}$ & $\gamma$ & $n$ & $p$ & $\nu$ & $\mu^{\pm} $ & $\pi^{\pm} $ }
    \startdata
            Injection & $Q_{e^-,\mathrm{inj}}$ & -- & $Q_{\gamma,\mathrm{inj}}$ & -- & $Q_{p,\mathrm{inj}}$ & -- & -- & -- \\ \hline 
            Escape & $\alpha_{e^-, \mathrm{esc}}$ & $\alpha_{e^+,\mathrm{esc}}$ & $\alpha_{\gamma,\mathrm{esc}}$& $\alpha_{n,\mathrm{esc}}$ & $ \alpha_{p,\mathrm{esc}}$
                        & $\alpha_{\nu,\mathrm{esc}}$ & $\alpha_{\mu,\mathrm{esc}}$ & $\alpha_{\pi,\mathrm{esc}}$ \\ \hline
            Synchrotron & $\dot{E}_{e^-,\mathrm{SY}}$ & $\dot{E}_{e^+,\mathrm{SY}}$& $\alpha_{\gamma,\mathrm{SY}}$, $ Q_{\gamma,\mathrm{SY}}$ &  -- & $ \dot{E}_{p,\mathrm{SY}}$ 
                       & -- & $ \dot{E}_{\mu,\mathrm{SY}}$ & $\dot{E}_{\pi,\mathrm{SY}}$ \\ \hline
            Inverse Compton& $ \dot{E}_{e^-,\mathrm{IC}}$ & $ \dot{E}_{e^+,\mathrm{IC}}$& $ \alpha_{\gamma,\mathrm{IC}}$, $ Q_{\gamma,\mathrm{IC}}$ & -- & $ \dot{E}_{p,\mathrm{IC}}$ 
                        & -- & $ \dot{E}_{\mu,\mathrm{IC}}$ & $\dot{E}_{\pi,\mathrm{IC}}$ \\ \hline
            Pair annihilation & $ Q_{e^-,\mathrm{pair}}$ & $Q_{e^+,\mathrm{pair}}$ & $\alpha_{\gamma,\mathrm{pair}}$ & -- &--  
                    &-- &-- & --\\\hline
            Bethe-Heitler & $Q_{e^-,\mathrm{BH}}$ & $ Q_{e^+,\mathrm{BH}}$ & -- & -- &  $ \dot{E}_{p,\mathrm{BH}}$
                    &-- & -- & -- \\\hline
            Photo-pion & -- & -- & $\alpha_{\gamma,\mathrm{p\gamma}}, Q_{\gamma,\mathrm{p\gamma}}$ & $\alpha_{n,\mathrm{p\gamma}},~Q_{n,\mathrm{p\gamma}}$ &  $\alpha_{p,\mathrm{p\gamma}},~Q_{p,\mathrm{p\gamma}}$ 
                        & -- & -- & $Q_{\pi,\mathrm{p\gamma}}$ \\ \hline
            Proton-proton & -- & -- & $Q_{\gamma,\mathrm{pp}}$ & -- &  $\dot{E}_{p,\mathrm{pp}}$ 
                        & -- & -- & $Q_{\pi,\mathrm{pp}}$ \\ \hline
            Adiabatic/Expansion&  $\dot{E}_{e^-,\mathrm{ad}}$, $\alpha_{e^-,\mathrm{exp}}$ &$ \dot{E}_{e^+,\mathrm{ad}}$, $ \alpha_{e^+,\mathrm{exp}}$  & $\alpha_{\gamma,\mathrm{exp}}$& $ \dot{E}_{p,\mathrm{ad}}$, $\alpha_{p,\mathrm{exp}}$ & $\alpha_{n,\mathrm{exp}}$
                    &$\alpha_{\nu,\mathrm{exp}}$ & $\dot{E}_{\mu,\mathrm{ad}}$, $\alpha_{\mu,\mathrm{exp}}$& $\dot{E}_{\pi,\mathrm{ad}}$, $\alpha_{\pi,\mathrm{exp}}$\\ \hline
            Pion Decay & --  & --  & --& -- & -- 
                    & $Q_{\nu,\mathrm{\pi-dec}}$& $Q_{\mu,\mathrm{\pi-dec}}$& $\alpha_{\pi,\mathrm{\pi-dec}}$ \\
            Muon Decay & $Q_{e^-,\mathrm{\mu-dec}}$ & $Q_{e^+,\mathrm{\mu-dec}}$ & --  & -- & --  
                    & $Q_{\nu,\mathrm{\mu-dec}}$ & $\alpha_{\mu,\mathrm{\mu-dec}}$ & -- \\ \hline
         \enddata
         \tablecomments{Neutral pions are not listed as a separate species here, as they decay instantaneously into photons. Instead, we list thus a source term for photons from photo-meson production $Q_{\gamma,\mathrm{p\gamma}}$ and $Q_{\gamma,\mathrm{pp}}$ .} 
\end{deluxetable}

\subsection{Solver overview}
\label{sec:solver}
The particle PDEs are solved on a logarithmic energy grid, using $x=\ln (E/E_0)$ with the energy scale $E_0$. While in the user interface, all energies are given in eV, internally $E_0=m_ec^2$ for electrons/positrons/photons and $E_0=1~\mathrm{GeV}$ for protons/neutrons/pions/muons/neutrinos
Then, Eq.~\ref{equ:overalllinearform} becomes 
\begin{equation}
	\partial_{t}n(x,t)=-\partial_{x}\left[A(x,t)n(x,t)\right]-\alpha(x,t)n(x,t)+\epsilon(x,t) \, ,
	\label{equ:overalllogarithmicform}
\end{equation}
with the substitutions 
\begin{equation}
	x=\ln \frac{E}{E_0}, \hspace{5mm} 
	n(x)\equiv\frac{\dd^2 N}{\dd x \dd V} = E n(E), \hspace{5mm}
	A(x)=\dfrac{\dot{E}(E)}{E}, \hspace{5mm}
	\epsilon(x)=E Q(E), \hspace{5mm}
	\alpha(x)=\alpha(E)\, .
\end{equation}
The x-axis is discretized along a finite-difference grid from $x_\mathrm{min}$ to $x_\mathrm{max}$ with equal spacing $\Delta{x}$. Particles at the $i$-th grid point (denoted as $n_i$) thus have logarithmic energy 
\begin{equation}
	x_{i}=x_\mathrm{min}+(i-1)\Delta{x}, \hspace{5mm}i=1,2,...i_\mathrm{max} \, .
\end{equation}
The energy grid spacing is fixed to $\Delta{x}=0.1$, which corresponds to approximately 23 energy bins per decade, which cannot be changed at the user level, as explained in Sec. \ref{sec:interface}. On the one hand, an alternative discretization technique like the finite volume method could offer more flexibility because it would not require fixing a structured grid and could, therefore, potentially allow for a more dynamic quantization of the integration axes. Nonetheless, our fixed finite difference grid offers a robust framework that has been extensively tested, and the prescribed spacing represents the optimal point of compromise between computational efficiency and precision in the calculation of the interaction kernels (cf. details in App.~\ref{app:solver}). Moreover, there is some flexibility regarding efficiency optimization in the form of user-adjustable values of $x_\mathrm{min}$ to $x_\mathrm{max}$, as explained in Sec. \ref{sec:interface}.

The time evolution in \AM is performed sequentially for the spectrum of each particle species by (1) updating the coefficients $A$, $\alpha$, and $\epsilon$ based on the current densities and parameters of the radiation zone, and then (2) using the solver algorithm to evolve the particle spectra by a small step $\Delta t$ in time. Both steps are performed sequentially for each species in the following order: protons, neutrons, pions, muons, neutrinos, electrons/positrons, and finally, photons.\footnote{Our discretisation choice in \AM is motivated by the order in which the particles are mainly produced.}

The equation system is stiff since the equations are coupled between different energy scales and particles. For instance, the synchrotron loss time scale is $t_\mathrm{syn} =E / \dot{E}_{\mathrm{SY}} \propto E^{-1}$. Hence, the dynamical range attributed to this process is many orders of magnitude (or equal to the energy grid size). Typical stability criteria, such as the Courant-Friedrichs-Lewy stability criterion, would force the time step $\Delta t$ to be driven by the smallest time scales (corresponding to the largest eigenvalue of the coefficient matrix). This leads to many time steps needed to calculate a steady state (or a quasi-steady state for slowly changing properties of the emission zone). \AM uses a combination of a tri-diagonal matrix solver with analytical approximations in different energy regimes as techniques to reduce the impact of stiffness. This allows the code to use larger time steps, resulting in high computational speed with only minor sacrifices to precision (cf. Sec.~\ref{sec:efficiency}).

\begin{figure}
    \centering
    \includegraphics[width=0.9\linewidth]{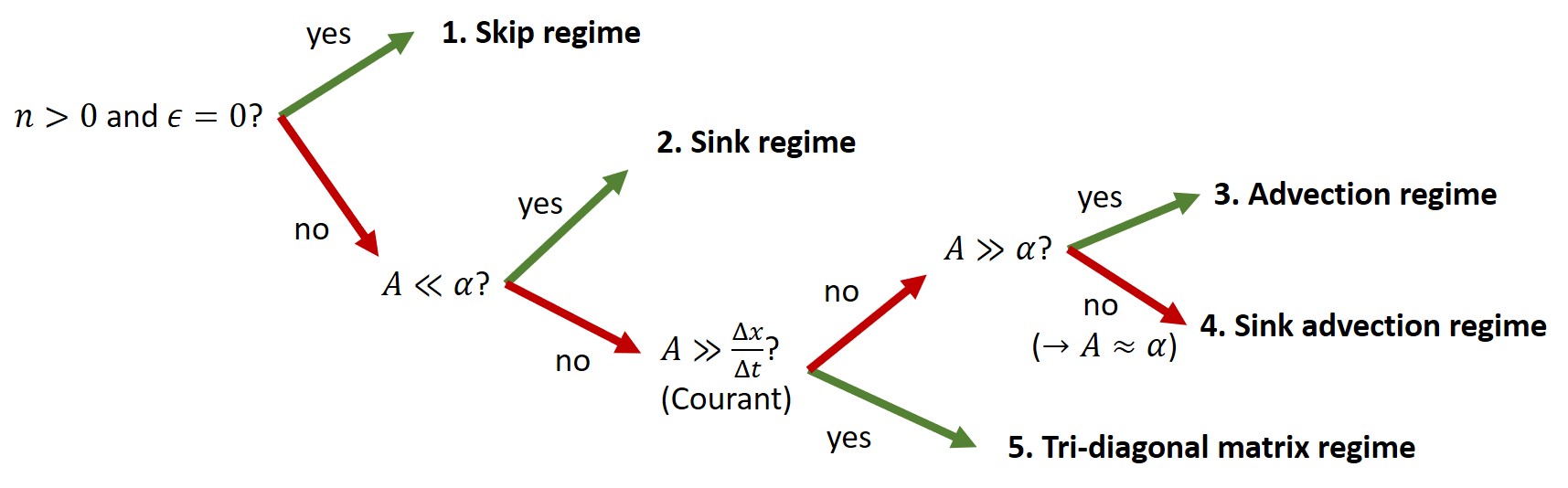}
    \caption{Illustration of the automatic process of selection of the solver algorithm: if both particle density and injection are zero, no action is required ("Skip" regime). If either sink/escape or advection dominate, analytical solutions neglecting the other term are applied ("Sink" and "Advection" regimes). If both are relevant, the PDEs are solved either through a tri-diagonal matrix algorithm ("Tri-diagonal matrix" regime),  or, if the tri-diagonal matrix method is unstable, through an analytical approach ("Sink-advection" regime).}
    \label{fig:solver}
\end{figure}

These different solver regimes are determined by comparing actual coefficients computed for the next time step against certain threshold criteria.
\AM distinguishes between the following solver regimes (see Fig.~\ref{fig:solver}):
\begin{enumerate}
    \item \textbf{``Skip" regime:} As the particle density and injection are zero [$n(x)=\epsilon(x) = 0$], no update of $n(x)$ is performed.
    \item \textbf{``Sink" regime:} Escape/sink terms are the dominant loss process [$\alpha(x) \gg A(x)$]. An analytic approximation is applied ($A=0$), approximating the current coefficients at $t$ to be constant until $t+ \Delta t$.
    \item \textbf{``Advection" regime:} Advection in energy space is the dominant loss process [$A(x) \gg \alpha(x)$]. An analytic approximation is applied ($\alpha=0$), approximating the current coefficients at $t$ to be constant until $t+ \Delta t$.
    \item \textbf{``Tri-diagonal matrix" regime:} Both escape/sink and advection are relevant, the tri-diagonal matrix method is applicable [$A(x) \approx \alpha(x), A(x) \ll \Delta x/\Delta t$]. It uses a semi-implicit Chang-Cooper discretisation scheme, i.e. weighted energy mid-points (as in \citet{1970:ChangCooper}) and averaged time mid-points (following the Crank-Nicolson scheme).
    \item \textbf{``Sink-advection" regime:} Both escape/sink and advection are relevant, but the tri-diagonal matrix method is unstable [$A (x) \approx \alpha(x), A(x) \gg \Delta x/\Delta t$]. An analytic approximation is applied, approximating the current coefficients at $t$ to be constant until $t+ \Delta t$.
\end{enumerate}
More details on regimes 2--5 and the implementation of the different methods can be found in App.~\ref{app:solver}. 
In principle, the correct solver method is chosen for each particle species at each energy. However, in order to avoid boundary effects when patching together solutions from different regimes at different energies we solve each advective regime (3, 4, 5) in a 10-bins larger energy range, where the extra bins are cut away during patching. This may create small features in the proton/pion/muon/electron/positron spectra, which however are typically washed out for observable secondaries (neutrinos and photons). We also highlight that for photons, neutrinos and neutrons no advection terms exist. This means that for those species, the explicit methods of regimes 1 or 2 are always used.

Finally, we discuss the choice of discrete time step $\Delta{t}$, which is inherent to the entire solver algorithm. In \AM, the value of $\Delta{t}$ can be adjusted at the user level. The selected value results in a trade-off between accuracy, stability, and computational cost. A simple but effective test of the stability of the results is the performance of simulations using gradually smaller time steps, and checking the consistency of the resulting particle and radiation spectra. It is also important to note that the analytical approximations build on the assumption that the coefficients $A$, $\alpha$ and $\epsilon$ (which sometimes depend in turn on the particle spectra) are constant during $\Delta t$. While this is often a good simplification, this assumption may be violated when there is strong feedback between different species, such as in the case of fast electromagnetic cascade development. This emphasizes the importance of checking the stability of the results, particularly when simulating extremely optically thick environments, and to adjust the solver time step accordingly. While this set up places considerable responsibility on the user, it also presents an opportunity for performance optimization in a flexible manner.

\subsection{User interface}
\label{sec:interface}
\AM can be accessed in its native language C\texttt{++} or through a Python~3.x interface. The Python interface provides `get' functions that allow access to the arrays containing the spectral densities of all particle species, the cooling times, and the dynamical quantities, at any point during runtime. Quantities representing parameters of the solver, physical parameters, particle densities, and particle injection rates can also be defined by the user by means of `set' functions. Below we provide a high-level overview of the quantities accessed and defined by the user in order to operate the software, which can be grouped into five categories. A detailed list of these functions ad their usage can be found in the software documentation\textsuperscript{\ref{footnote:source_code}}.

\subsubsection*{Energy grids}

As explained in Sec.~\ref{sec:solver}, the particle density spectra and interaction rates are calculated along discretized energy grids. For optimization purposes, the code works with four distinct energy grids: one for protons, neutrons, pions, and muons; one for electrons and positrons; one for photons; and one for neutrinos. The user can retrieve these four energy grids in the form of arrays, by means of four dedicated get functions. At the Python level, the user can neither arbitrarily set each of these grids, nor should this be done in the source code without careful consideration of the implications on numerical stability. Instead, for the electron and positron grid the minimum energy is set to the electron rest mass energy $m_e c^2$, whereas for the protons, neutrons, pions and muons the minimum energy is set to $1\,$GeV.
Furthermore, the user can specify a common maximum energy for all particle grids, the minimum energy of the photon grid, and the minimum energy of the neutrino grid. 
Internally, this user input is accommodated through the resizing of the four above-mentioned energy grids. There are two limitations due to the intrinsic nature of the numerical solver: \textit{1)} the individual energy bin size is fixed to $\Delta(\ln E)=0.1$ for all particles, as explained in Sec. \ref{sec:solver}, and \textit{2)} the energy grids cannot be changed by the user during the simulation, but only before initializing the solver.

\subsubsection*{Parameters characterizing the radiation zone}

The radiation zone is fully characterized\footnote{Additional physical properties that do not affect the interactions in the comoving frame of the radiation zone, such as the source redshift and the bulk Lorentz factor of the radiation zone, can be included a posteriori, e.g., by applying the respective Doppler boost factor to the emitted radiation.} by the strength of the homogeneous magnetic field (relevant for synchrotron emission, absorption, and cooling), the escape timescale (used to specify the escape rates for each particle species) and the expansion timescale (relevant for adiabatic cooling and density dilution due to the plasma expansion). All parameters may be adjusted at each time step by means of `set' functions. Updating the magnetic field strength automatically triggers a re-computation of the synchrotron kernels. 
Since the computation depends only on the particle densities (in the comoving frame), no assumptions on the zone's geometry or expansion are included a priori in the software; instead, they are encoded in the escape and expansion timescales provided by the user.
Finally, a cold proton target density can be specified which is used for proton-proton interactions.

\subsubsection*{Switches for physical processes and optimization routines}

Each of the interaction processes introduced above, as well as the respective performance optimization options described in App. \ref{app:physics_processes}, may be turned on or off by means of individual switches. Each of these switches is a dedicated function that accepts a binary input. For each process, it is also possible to switch on and off emission and cooling separately. For example, by switching off Bethe-Heitler emission and switching on Bethe-Heitler cooling, the user would be effectively removing from equation system the terms responsible for the emission of Bethe-Heitler pairs ($Q_{e^-,\mathrm{BH}}$ and $Q_{e^+,\mathrm{BH}}$) , while still including the radiative cooling effect of this process on the protons ($ \dot{E}_{p,\mathrm{BH}}$). This feature can be used to isolate given processes when carrying out a physics cross-check, or for optimization purposes, for example if it is known that a given process can be safely neglected.

\subsubsection*{Particle injection}

The user can define the particle density injection rate for any type of particle, in the form of numerical arrays provided as input to dedicated `set' functions. For protons and electrons, there are also built-in functions for the injection of template spectra, namely simple or broken power laws. In that case, rather than providing an array containing the energy-dependent injection rates, the user provides the power-law spectral indices, the maximum, minimum, and break energies, the total injection luminosity, and the volume for the calculation of the density. In either case, setting the injection introduces a source term $Q(E)$ for the respective species (cf. Sec.~\ref{sec:equation_system}). This term remains constant in time until a new injection is defined.

\subsubsection*{Particle densities}

The number density spectrum $n(E,t)$ of each particle species (cf. Sec.~\ref{sec:equation_system}) is stored at each timestep in the form of an array whose size is that of the corresponding energy grid. Upon initialization, all particle types have a density of zero across the spectrum. At any time step, the user can consult and redefine the density array of each particle type, using dedicated `get' and `set' functions, respectively. Defining a given density spectrum may be useful, for instance, for setting up the initial conditions, like in the case of external photons that are already present in the system at the beginning of the simulation, as in the example discussed in Sec.~\ref{sec:applications_blazar}.

It is also possible to access the density spectra of particles of a given species produced through a specific interaction channel. For example, it is possible to consult separately the density spectrum of pairs produced through Bethe-Heitler emission and those produced in the decay chain of photo-pions. Similarly, it is possible to consult separately the density spectrum of photons produced through synchrotron emission (or inverse Compton scattering\footnote{This would be defined as the photon spectrum produced by a given individual sub-population of electron-positron pairs (e.g., originating only from the Bethe-Heitler process) when inverse Compton scattering with the full photon distribution.}) by each of these distinct electron populations. To utilize this ``particle-tracing'' feature, the user must turn on a dedicated switch before the simulation initialization; to improve performance, this feature should be kept off since it costs the solver additional computation time for maintaining additional species that are copies of the original particle species but with only a single source term active. For instance, tracing the pairs from Bethe-Heitler production involves a species with all the loss and sink terms as normal electrons and positrons but whose only source terms are $Q_{\mathrm{e}^\pm, BH}$. This feature helps trace the physical processes that give rise to certain features in the particle spectra and calculate the detailed break-downs into individual components as in Figs.~\ref{fig:blazar_plots},~\ref{fig:TDE_multimessenger}~and~\ref{fig:GRB_example}.

\subsubsection*{Interaction timescales}

The interaction timescale of each physical process, which can essentially be derived from the corresponding coefficient in Tab.~\ref{tab:coefficients}, can be read out at every time step by means of `get' functions. For each particle species and each physical process, there is a dedicated get function that returns an array of energy-dependent timescales; the size of this array is that of the energy grid of the respective particle species. Unlike particle density spectra, the interaction timescales cannot be `set' directly by the user, but instead result from the internal time-dependent computation.

\section{Applications}
\label{sec:applications}

We now discuss examples of three astrophysical source classes modeled in recent works using \AM. We first introduce a blazar study, providing extensive details on the photon emission from each individual radiation process and the respective cooling timescales. We then discuss a TDE model and a GRB model from the recent literature, as examples of 
different astrophysical applications of the software. 

Other scenarios have also been tested in the context of the Hadronic Code Comparison Project \citep{Cerruti:2021hah}, including non-linear time-dependent scenarios such as the inverse Compton catastrophe and the  Bethe-Heitler pair production loop. In all cases, \citet{Cerruti:2021hah} have found the \AM results compatible with those from other lepto-hadronic softwares (with deviations below $\pm 10$~\% for leptonic, and below $\pm 40$~\% for lepto-hadronic scenarios).

\subsection{Example of a steady-state model: blazar~PKS~0736+01}
\label{sec:applications_blazar}

Blazar PKS~0736+01 is a flat-spectrum radio quasar (FSRQ) lying at redshift 0.19 \citep{2009ApJS..184..398H}, with a $\gamma$-ray luminosity of $6\times10^{45}~\rm erg/s$ as detected by the \textit{Fermi} Large Area Telescope (\textit{Fermi}-LAT). The source was also recently detected at energies above 100~GeV by the H.E.S.S. telescope~\citep{2020A&A...633A.162H}, making it the seventh FSRQ at that time to be detected in very-high-energy $\gamma$-rays.

As explained in more detail in \citet{Rodrigues:2023vbv}, the multi-wavelength broadband emission from PKS~0736+01 can be described by a one-zone lepto-hadronic model. In that study, the bulk of the emission in the optical and $\gamma$-ray bands was assumed to originate in leptonic interactions in all sources, while in some sources the X-ray spectrum was shown to be explained with an additional component from hadronic cascades, as was demonstrated to be the case of PKS~0736+01. The neutrino flux originating in the decay chain of photo-pions was then self-consistently estimated for each source.

In the relativistic jet, electrons and protons are assumed to be accelerated to power-law spectra. The acceleration mechanism is not explicitly simulated in \AM; instead, electrons and protons are injected, as described in Secs.~\ref{sec:interface}, with a rate that is constant in time, following power-law spectra. The dissipation region (i.e. the zone in the jet where the radiative interactions take place, which is the zone simulated with the code) is assumed to be spherical, with a radius $R_\mathrm{b}^\prime$, and permeated with a homogeneous and isotropic magnetic field of strength $B^\prime$. Tab~\ref{tab:blazar_parameters} lists the best-fit parameter values of the one-zone model.

\begin{table}[H]
    \caption{Simulation parameters for the blazar PKS~0736+01 discussed in this section. The primed symbols represent the values in the rest frame of the relativistic jet (i.e. the radiation zone), which is the reference frame of our \AM simulation. Other AGN parameters, such as the bulk Lorentz factor of the jet, can additionally be found in Tab.~B.1. of \citet{Rodrigues:2023vbv}. These do not play a role in the numerical simulation itself, but only in converting these quantities into the observer's frame.}
    \centering
    \begin{tabular}{lllllllllll}
    \hline
    $R_\mathrm{b}^\prime$ [cm] & $B^\prime$ [G] & $L_\mathrm{p}^\prime$ [erg/s] & $L_\mathrm{e}^\prime$ [erg/s] & 
    $\gamma_\mathrm{p}^{\prime\mathrm{min}}$ &
    $\gamma_\mathrm{p}^{\prime\mathrm{max}}$ &
    $\gamma_\mathrm{e}^{\prime\mathrm{min}}$ & 
    $\gamma_\mathrm{e}^{\prime\mathrm{max}}$ & 
    $\alpha_\mathrm{p}$ &
    $\alpha_\mathrm{e}$ & $\Gamma_\mathrm{bulk}$\\
    \hline
    $5.0\times10^{16}$ & $7.9\times10^{-1}$ &
    $6.3\times10^{43}$ & $2.0\times10^{41}$ & 
    $1.0\times10^2$ & $2.0\times10^{6}$&
    $2.0\times10^{01}$ & $7.9\times10^{3}$ & 1.0 & 2.0 & 17.6\\
    \hline
    \end{tabular}    \label{tab:blazar_parameters}
\end{table}

\begin{figure}[tb]

    \includegraphics[width=\linewidth]{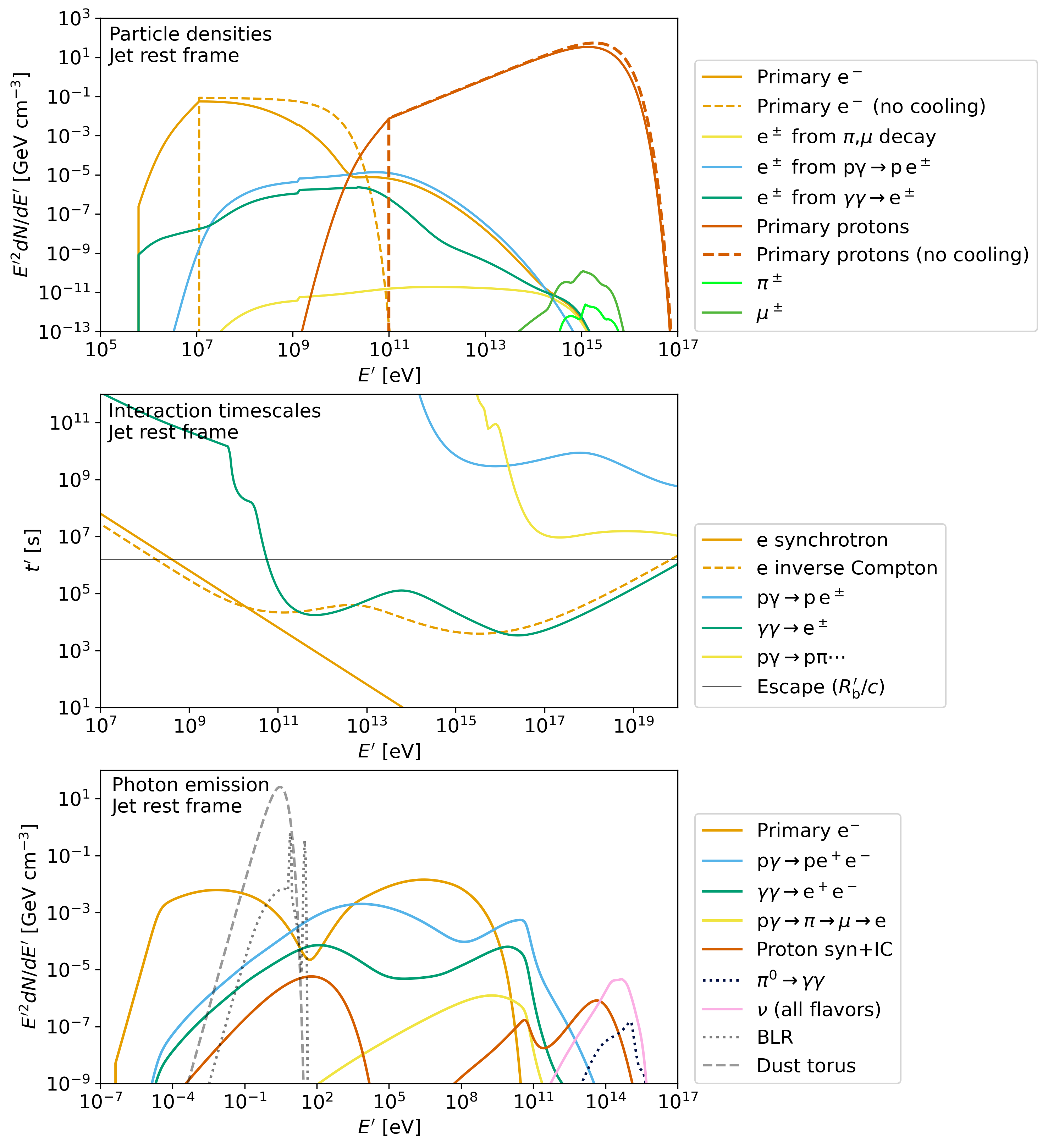}
        
    \caption{Results of the \AM simulation of the flat-spectrum radio quasar PKS~0736+01~\citep{Rodrigues:2023vbv}. \textit{Top:} Steady-state energy density spectra of charged particles in the rest frame of the jet. The dashed light orange and dark orange curves represent the injection spectrum of electrons and protons, while solid curves represent steady-state spectra. \textit{Middle:} Interaction timescales in the steady state. The gray line shows the light-crossing time of the zone, which represents the escape timescale. The values of $\tau$ represent the optical thickness of synchrotron emission at maximum electron energy; photon annihilation at 1~TeV; and Bethe-Heitler pair production at the maximum injected proton energy. \textit{Bottom:} Resulting steady-state photon energy densities. The different colors represent the different processes from which the photons originate, as described in the legend. The pink curve shows the all-flavor steady-state neutrino spectrum. The gray dashed curve represents the external photon fields that are relativistically boosted into the rest frame of the jet from the surrounding BLR \citep[for details, see][]{Rodrigues:2023vbv}.}

    \label{fig:blazar_plots}

\end{figure}

To describe the observed fluxes in a lepto-hadronic framework, the dissipation zone should be lying in relative proximity 
to the broad line region (BLR). The BLR, a region of dense and rapidly rotating gas surrounding the supermassive black hole, emits broad lines as a consequence of being illuminated by the accretion disk. 
As determined by means of multi-wavelength data fitting \citep{Rodrigues:2023vbv}, in this source the dissipation region should lie at a distance of $R_\mathrm{diss}=2.5\,R_\mathrm{BLR}$ to the supermassive black hole, where $R_\mathrm{BLR}$ is the radius of the BLR, assumed here to be a thin spherical shell. That allows us to convert the luminosity of the broad line emission in the black hole rest frame to a photon density in the rest frame of the dissipation region, following the procedure detailed in \citet{Rodrigues:2023vbv}.

Fig.~\ref{fig:blazar_plots} summarizes the simulation results. In the upper panel, we show as dashed curves the energy density spectrum of injected electrons (light orange) and protons (dark orange). These are assumed to be simple power laws with an exponential cutoff, whose parameter values are detailed in Tab.~\ref{tab:blazar_parameters}. As the simulation proceeds, the accelerated cosmic ray protons and electrons undergo energy losses. As we can see in the middle panel of Fig.~\ref{fig:blazar_plots}, in this case electrons cool efficiently through both synchrotron and inverse Compton emission (orange line and dashed curve). At the maximum energy, the optical thickness to these processes is of the order of $\tau\sim10$, as indicated in the figure. This optical thickness is given by the ratio between the cooling time and the escape time, which we assume to be given by the light-crossing time of the region, $R_\mathrm{b}^\prime/c$ (note that adiabatic cooling is not included). This represents a scenario where all particles escape the region advectively, for instance due to relativistic winds.

The effect of cooling on electrons can be seen in the upper panel, as the steady-state electron spectrum (solid orange curve) displays a break compared to the injected ones. On the other hand, the steady-state proton spectrum does not considerably change compared to the injected power law, because the system is optically thin to proton interactions (cf. blue curve in the middle panel of Fig.~\ref{fig:blazar_plots}).

The lower panel of Fig.~\ref{fig:blazar_plots} shows the emitted photon spectra, given as energy density spectra in the jet rest frame. The dashed gray curve represents the external photons from the BLR, described above: the two peaky features in the external photon spectrum originate in broad line emission from the hydrogen and helium gas present in the BLR, while the thermal bump represents the fraction of the accretion disc emission that gets isotropized in that region through Thomson scattering. Unlike the components shown as solid curves, the photon fields from the BLR are not produced self-consistently in the course of the simulation, because they represent an external source of radiation that has to be added manually as part of the model-building process. Computationally, this is accomplished with a three-step process: first, the photon density spectra are calculated and interpolated on the \AM photon energy array (cf.~Sec. \ref{sec:interface}). After the simulation is initialized, the photon densities $n(E)$ (which by default are zero at the beginning of the simulation) are set to the calculated density values, which represent the fact that the external radiation is present in the simulation environment since the beginning, with a constant density. In our case, the initial photon density does not affect the results, because we focus on the steady-state emission; however, this procedure is a robust and general way to represent the simulated environment from the initial time steps. Finally, we set a constant injection term  with the same spectral shape, which guarantees that the BLR photons are constantly injected at every time step. Upon calculating this spectrum as shown in the figure, it is injected in \AM as an external species, in the same way as the accelerated protons and electrons. As detailed previously in this work, this effectively means that the software includes a constant injection term for these three species at every time step throughout the duration of the simulation.

In the beginning of the simulation, the magnetic field and the external photons are the only targets for electron and proton interactions. As the solver evolves, synchrotron and inverse Compton emission lead to significant photon fluxes, shown in their steady state as a light orange curve in the lower panel of Fig.~\ref{fig:blazar_plots}). Protons also interact, in this case primarily through the Bethe-Heitler process. As shown by the blue curve in the upper plot, the spectrum of these pairs extends up to $10~\mathrm{TeV}$ energies, much above the primary electrons. At these energies, synchrotron emission is extremely efficient, resulting in the large fluxes shown in blue in the lower plot. It is interesting to observe that this flux level is comparable to that from primary electrons, in spite of the source being optically thin to Bethe-Heitler process ($\tau\sim10^{-4}$ as shown in the timescale plot). This is because 1) we inject a much larger power in primary protons compared to electrons ($L_\mathrm{p}^\prime=3\times10^3\,L_\mathrm{e}^\prime$) and 2) Bethe-Heitler pairs are produced at much higher energies compared to primary electrons, a regime where their emission is more efficient by several orders of magnitude.   

Returning to the timescale plot, we can see from the green curve that photons above a few tens of GeV are efficiently attenuated through photon-photon annihilation, reaching an optical thickness of 100 at TeV energies.  This process is responsible for the strong attenuation of the photon spectrum at TeV energies that can be seen in the lower panel. It also leads to the emission of pairs with energies of up to PeV energies (green curve in the upper plot). The emission from these pairs leads to the photon spectrum shown in green in the lower panel, which in this case is subdominant compared to the emission from Bethe-Heitler pairs. We can see two suppression features in this curve: the one at $\sim10^{12}~\rm eV$ is due to attenuation on the external photon fields, while the one at $\gtrsim10^{16}~\rm PeV$ is mainly due to attenuation on the synchrotron photons from the primary electrons.

Finally, we discuss the photo-meson process, responsible for the emission of high-energy neutrinos. This particular source is optically thin to this process, as we can see by the yellow curve in the middle panel. This can be also seen by the ratio between the steady-state luminosity in protons and that in pions (and muons), shown in green (yellow) in the upper panel. 
Neutrinos (pink curve in the bottom plot) follow the injection spectrum of pions and muons. The latter may be estimated by switching off pion and muon decay in the calculations (dashed curves in the upper plot)
On the other hand, the electrons that are produced in the decay of muons and pions quickly cool due to synchrotron emission, leading to the very soft spectrum shown in light orange in the upper panel. It is interesting to consider that under the same conditions, if protons were accelerated up to energies above 100~PeV (in the comoving frame) the optical thickness to photo-meson production would increase by a factor of $\sim10^5$, as can be seen from the yellow curve in the middle panel. This is due to the interaction with the external photons from the broad line region, which becomes viable above these energies.

It is the combined modeling of all these species over a broad energy range that enables a self-consistent description of the emerging photon and neutrino spectra, while the ability to read out the timescales and track the contributions of the various processes allows for a deeper understanding. Further, the self-consistent calculation of secondary particle spectra allows to track how energy is transferred from high to lower energies, e.g., through lepto-hadronic photo-pair production, where in consequence additional components at energies of $\mathcal{O}$(100~eV) may constrain neutrino data.

\subsection[]{Time-dependent source modeling: TDE example\footnote{ A detailed example of the time-dependent multi-messenger modeling of a neutrino-coincident TDE, based on \cite{2024arXiv240109320Y}, is available at \url{https://am3.readthedocs.io/en/latest/TDE_example.html}}}
\label{sec:applications_tde}

In recent years, growing evidence is pointing towards TDEs (where a star is tidally disrupted by a supermassive black hole) as sources of high-energy neutrinos \citep[e.g.,][]{2021NatAs...5..510S, 2022PhRvL.128v1101R, 2021arXiv211109391V}. One common feature of these neutrino-coincident TDEs is that the neutrino events are detected months after the peak of optical and ultraviolet (OUV) light curves. The neutrino counterparts together with the multi-wavelength observations, including OUV/infrared (IR)/X-ray detection and \emph{Fermi}-LAT/HAWC upper limits in $\gamma$-ray bands, may shed light on the intrinsic TDE properties.
 In contrast to AGNs, TDEs are transient sources for which a steady-state assumption may not adequately describe the particle distributions over a time window of hundreds of days, as the particle cooling and escape times are not always shorter than the variation time of proton injection \citep{Yuan_2023} This implies the proton source term $Q(E, t)$ should be updated over time. Moreover, for obtaining reliable results, a fine-tuned time step should be used to prevent exceeding the interaction timescale of the most efficient processes. This is particularly important in cases involving dense target photon fields or strong magnetic fields.

\begin{figure*}[t]
    \centering
    \includegraphics[width=0.49\textwidth]{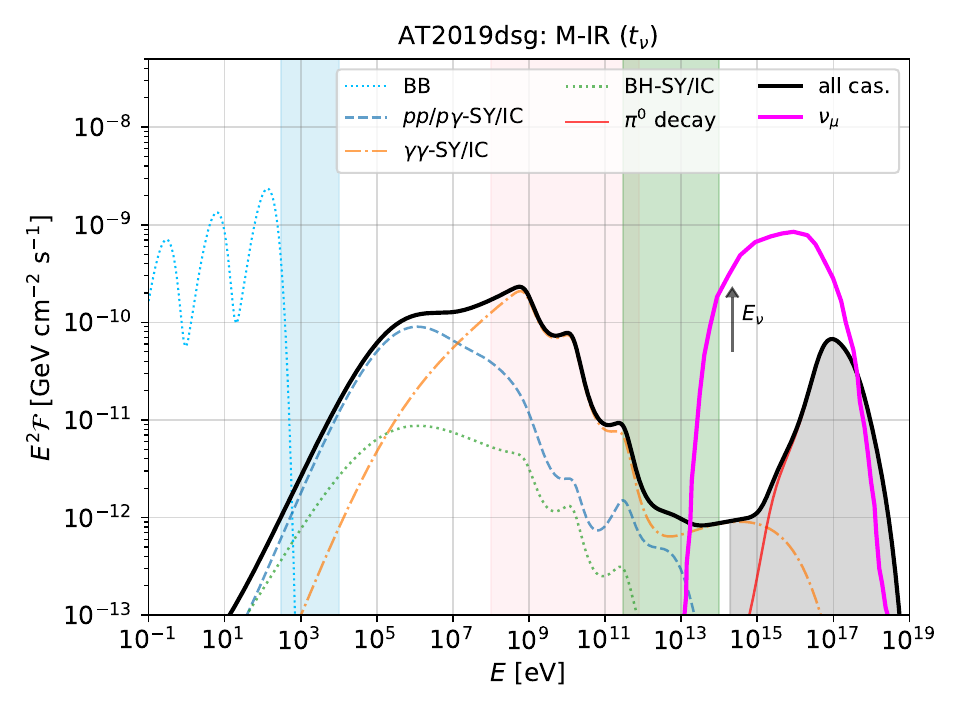}
    \includegraphics[width=0.49\textwidth]{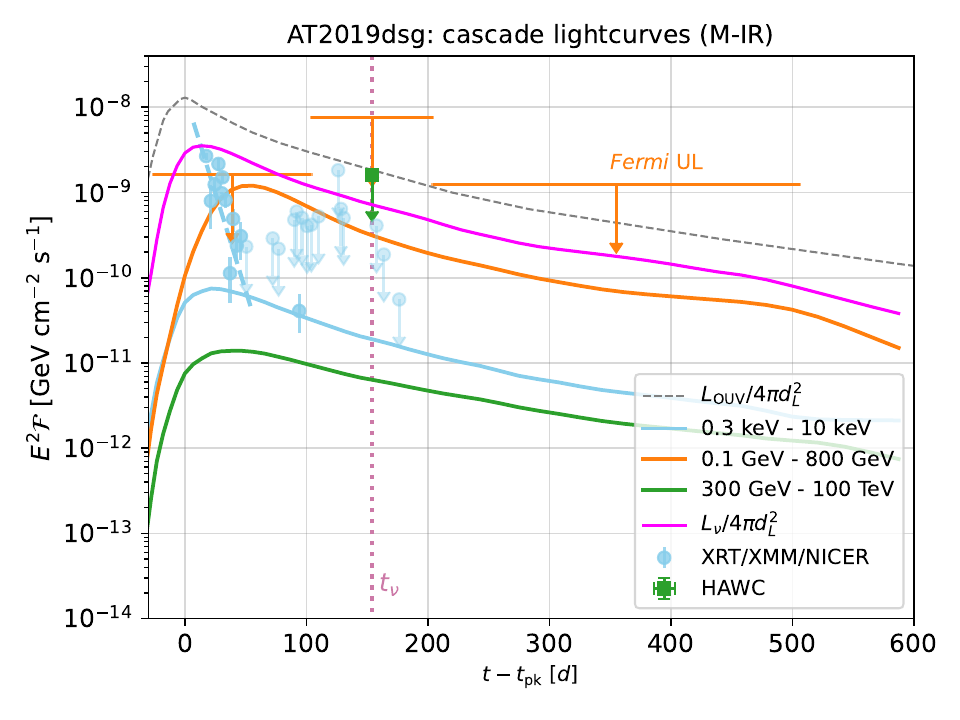}
    \caption{Time-dependent multimessenger modeling of TDE AT2019dsg. {\bf Left panel:} Electromagnetic cascade  and neutrino spectra at the time of the neutrino observation $t_\nu$. The black and magenta curves show the spectra of EM cascade and single-flavor neutrino emission. The black body IR, OUV and X-ray spectra are depicted as the thin cyan dotted curves. The thin blue dashed, orange dash-dotted, green dotted, and red curves represent the components contributing to the EM cascade. The blue, red, and green shaded areas depict the energy ranges of \emph{Swift}-XRT, \emph{Fermi}-LAT, and HAWC observations, whereas the gray-shaded areas indicate where $\gamma\gamma$ attenuation with the cosmic microwave background and extragalactic background light (EBL) prohibits the observation of the emitted $\gamma$-rays.
{\bf Right panel:}  Neutrino, $\gamma$-ray and X-ray light curves. The magenta, green, orange, and blue solid curves illustrate the single-flavor neutrino ($L_\nu/4\pi d_L^2$), very-high-energy (VHE) $\gamma$-ray (0.3 - 100 TeV), $\gamma$-ray (0.1 - 800 GeV), and X-ray light curves, respectively. The flux up limits from HAWC, \emph{Fermi}-LAT, \emph{Swift}-XRT/XMM/NICER and eROSITA \citep{2021NatAs...5..510S,2021MNRAS.504..792C,2021A&A...647A...1P,2022PhRvL.128v1101R} are shown as the green, orange, and blue points, respectively. Furthermore, the neutrino detection times are marked as the vertical magenta dotted curves. The OUV fluxes are also plotted as the gray dashed curves for reference. Note: the figures shown here are taken from \cite{Yuan_2023}, where a more detailed description can be found.}
    \label{fig:TDE_multimessenger}
\end{figure*}

Recently, \AM has been employed to investigate the signatures of electromagnetic cascade emission initiated by the photomeson interactions between thermal photons and energetic protons in TDEs \citep{Yuan_2023,2024arXiv240109320Y}. The model describes the emission from isotropic sub-relativistic winds, where the proton injection luminosity follows the OUV light curve, e.g., $L_{p}\propto t^{-5/3}$.  Noting that the TDEs are typically super-Eddington accretors, we normalize the peak proton injection luminosity to be a fraction of the peak accretion luminosity, e.g., $L_{p,\rm pk}=\epsilon_p L_{\rm acc,pk}$, where $L_{\rm acc,pk}$  is the peak accretion luminosity and could reach hundreds of Eddington luminosity. Motivated by the observations of thermal IR, OUV and X-ray emissions, a spherically symmetric radiation zone where the thermal photons are isotropized is considered. The radius of the radius zone is typical in the range $10^{15}-10^{17}$ cm \citep{2023ApJ...948...42W,Yuan_2023}. In the radiation zones with the presence of abundant thermal infrared, OUV and X-ray photons and magnetic field strengths typical of AGN jets, $B\simeq0.1-1~\rm G$, photo-meson ($p\gamma$) interactions can efficiently produce neutrinos and initiate EM cascades via $\pi^0$ decay and charged secondary particles such as $\pi^\pm$, $\mu^\pm$ and $e^\pm$. In addition, $e^\pm$ pairs originating from $\gamma\gamma$ attenuation and Bethe-Heitler (BH) processes could redistribute energies to the radiation field via synchrotron and inverse Compton radiation. 

\citet{Yuan_2023} have demonstrated the applicability of \AM to TDEs. The authors studied the spectral distribution and temporal evolution of the EM cascades for AT2019dsg and AT2019fdr for IR photon dominant (M-IR) and OUV photon dominant (M-OUV) scenarios \citep{2023ApJ...948...42W,Yuan_2023}.
Here, we introduce the M-IR scenario of AT2019dsg as an example to discuss the treatment of EM cascades in TDEs using \AM.  In this case, the radius of the radiation zone extends to the radius of the dust torus where the IR photons are generated via the dust sublimation, e.g., $R_{\rm IR}=5\times10^{16}~\rm cm$, and the injected proton maximum energy can reach $E_{p,\rm max}=5\times10^{9}$ GeV. The fiducial parameters $\epsilon_p=0.2$ and $B=1$ G are used. Given the proton injection rate, the distribution of target photons and the magnetic fields, \AM was employed to solve the time-dependent transport equations for all relevant particle species. The left panel of Fig.~\ref{fig:TDE_multimessenger} shows the SEDs for EM cascade (black curve) and neutrino (magenta curve) emission at the neutrino arrival time, approximately 154 days after the peak time ($t_{\rm pk}$) of OUV light curve. In this figure, the blue dashed, orange dashed-dotted, green dotted and red solid curves respectively represent the contribution of {$p\gamma\to\pi^\pm \to\mu^\pm \to e^\pm$, $e^{\rm }$ pairs from $\gamma\gamma\to e^\pm$ attenuation, Bethe-Heitler pair production ($p\gamma \to pe^\pm$) and $\pi^0\to\gamma\gamma$ dacay}. The peak of each component and the $\gamma\gamma$ absorption wiggles in the GeV - TeV range are consistent with analytical estimations \citep[a detailed description of SED shapes can be found in][]{Yuan_2023}.
We also show the spectrum of target thermal photons as blue dotted curve. The blue, pink and green areas depict the energy bands for {\textit{Swift}-XRT (0.3-10~keV), \textit{Fermi}-LAT} (0.1-800~GeV) and HAWC (300~GeV - 800~TeV). The light curves for the corresponding energy bands are demonstrated in the right panel of Fig.~\ref{fig:TDE_multimessenger}. The orange and green data points are the energy-integrated upper limits obtained by \textit{Fermi}-LAT and HAWC, whereas \emph{Swift}-XRT, XMM, and NICER observations are plotted as blue points. The \textit{Fermi}-LAT $\gamma$-ray up limits are obtained in three time slots, $t-t_{\rm pk}\sim-20-100{~\rm d},~{100-200~\rm d}$ and $200-500~\rm d$.

Remarkably, the EM cascade and neutrino light curves exhibit distinct temporal features comparing to the thermal OUV bolometric light curve, e.g., $L_{\rm OUV}/4\pi d_L^2$, as the black dashed curve in the right panel. We observe that the peaks of EM cascade light curves appear roughly $30-50$ days after the OUV peak, which is commonly referred to as the time delay in the TDE community. In the M-IR scenario, where the radiation zone is calorimetric ($p\gamma$ efficient) but optically thin to $p\gamma$ interactions, the time delay observed in the EM cascade emission can be interpreted as the $p\gamma$ timescale, e.g., $t_{p\gamma}\sim30-50$ days. 

Above we introduced the M-IR scenario for AT2019dsg as an illustrative example of employing \AM to model the time-dependent multi-messenger counterparts originating from continuously fueled TDEs. Given that the proton injection rate can undergo significant decrease over the photo-pion interaction timescale, which ranges from $10^{5}$ to $10^{7}$ seconds, the application of a steady-state approximation, by letting $\partial_t n(E,t)=0$ in Eq. \ref{equ:overalllinearform}, may prove inadequate for accurately portraying temporal characteristics, including time delays. 
In this case, the fully time-dependent treatment for the lepto-hadronic processes in \AM is thus crucial to study the spectral and temporal signatures of transient, though long-lasting and persistently-powered sources in the era of multi-messenger astrophysics.

\subsection{GRB modeling}
\label{sec:applications_grb}

GRBs have long been hypothesized to be a source of ultra-high-energy cosmic rays (UHECRs) and high-energy neutrinos \citep[e.g.,][]{Waxman:1995vg, Waxman:1997ti, Vietri:1998nm}. However, to date, no statistical association between measured high-energy neutrinos and the prompt phase of known GRBs could be made \citep{IceCube:2012qza, Aartsen:2014aqy,Aartsen:2017wea, IceCube:2022rlk}.
This implies that lepto-hadronic models of the prompt phase of GRBs have to obey neutrino limits. 
Within this context, \AM has been used to study the prompt phase of energetic GRBs as potential sources of UHECRs in \citet{Rudolph:2022ppp}, focusing on the maximal baryonic loading compatible with multi-wavelength and neutrino constraints. 
The short computation times of \AM allowed to implement a multi-zone internal shock model following \citet{Daigne:1998xc} where the unsteady GRB jet is approximated by $\mathcal{O}$(1000) plasma shells. These undergo $\mathcal{O}$(1000) collisions, in which particles are accelerated at collisionless shocks and participate in radiative processes. This simplified outflow model enables to separate different particle production regions along the jet, which is especially important in the context of neutrino predictions \citep{Bustamante:2014oka, Bustamante:2016wpu}.

In contrast to AGNs, GRBs are highly variable transient sources, where particle populations are unlikely to reach a steady state and are strongly impacted by non-linear effects.
As a consequence, they are typically modelled following the evolution over approximately one dynamical timescale. 
However, the high densities and strong magnetic fields of up to $\mathcal{O}$($10^6$~G) imply a complexity in the computations. For example, the latter make it necessary to invoke a quantum synchrotron treatment, as well as a full inclusion of secondary pion and muon cooling prior to their decay.

\begin{figure}[t]
    \centering
    \includegraphics[width=0.45\linewidth]{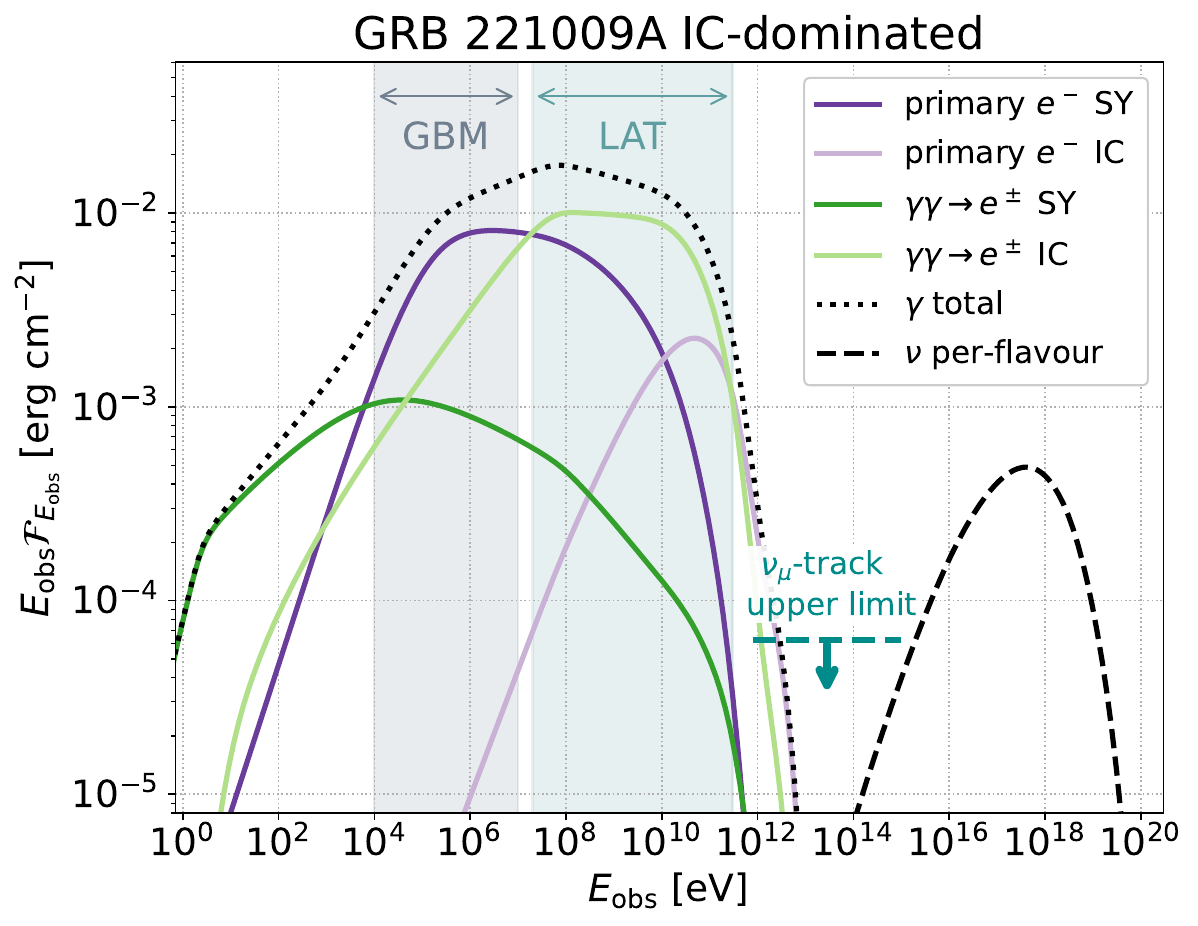}
    \includegraphics[width=0.45\linewidth]{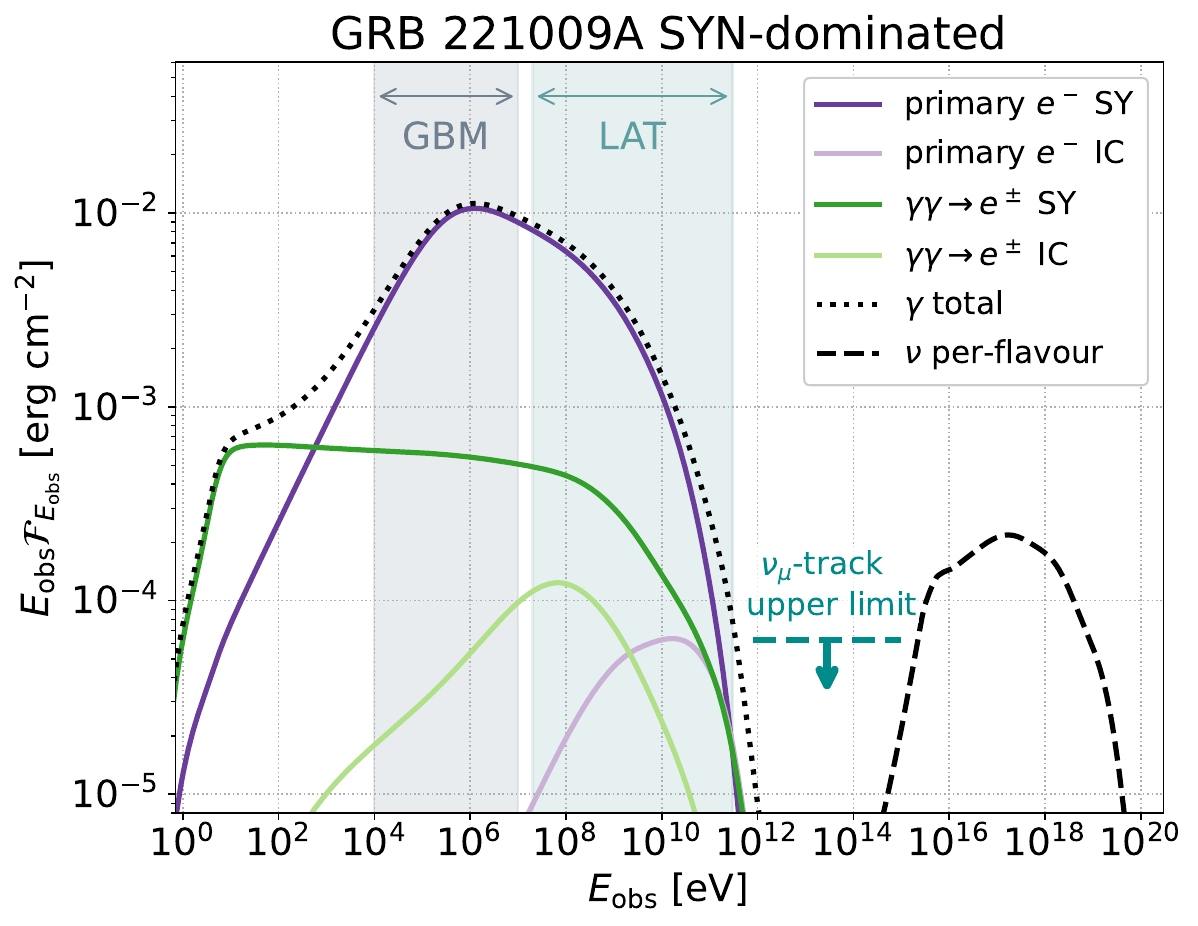}
    \caption{Time-integrated fluences for a synthetic GRB resembling GRB~221009A. Quantities are shown in the observer's frame, for a redshift of $z = 0.151$ and including EBL absorption using the model of \citet{Dominguez:2010bv}. The plots reproduce the "R16" scenario of \citet{Rudolph:2022dky}, split into the SYN-dominated scenario of strong magnetic fields on the left and the IC-dominated scenario of weak magnetic fields on the right. Due to internal and EBL absorption, direct signatures of hadronic processes are not present in the observed SED, which is instead dominated by synchrotron and inverse Compton emissions of primary electrons and lepton pairs produced in $\gamma \gamma$-annihilation. For neutrinos, we show the per-flavour fluence, that may be compared to the $\nu_\nu$-track upper limit reported in \citet{GRB221009A_IceCube_GCN32665}.}
    \label{fig:GRB_example}
\end{figure}
As a GRB application example, we show two synthetic time-integrated spectra in the observer's frame reproducing the properties of GRB~221009A in Fig.~\ref{fig:GRB_example}. 
GRB~221009A is one of the most energetic events detected, located at a relatively low redshift of $z = 0.151$. 
For the full details, we point the interested reader to \citet{Rudolph:2022dky}, where the examples shown here correspond to the ``R16" case. The left and right panel of the Figure depict a Synchrotron (``SYN") and inverse Compton (``IC") dominated scenario, respectively. While normalizing to the same emitted isotropic energy $E_\mathrm{iso}$ and keV peak energy of $\sim 1$~MeV similar to those reported for GRB~221009A, these scenarios are obtained by adjusting the fraction of energy transferred to magnetic field. Broadly speaking, the SYN-dominated scenario is realized for strong magnetic fields, while the IC-dominated scenario is realized for weak magnetic fields. Besides the total photon fluence (dotted lines), we indicate the contributions of primary electrons as well as lepton pairs produced in internal $\gamma \gamma$-annihilation, which dominate the SED. To guide the eye, we further indicate the approximaate energy ranges of \textit{Fermi} GRB Monitor (GBM, 10~keV - 10~MeV) and \textit{Fermi}-LAT (20~MeV - 300~GeV) and mark the reported $\nu_\mu$-track upper limit of IceCube \citep{GRB221009A_IceCube_GCN32665}.

Without going into detail, we highlight two features of the results. \\
\textit{(1) Importance of the cascade emission.}
Even though the event is located at a relatively small redshift $z = 0.151$, extragalactic background light (EBL) absorption (here calculated as in \citet{Dominguez:2010bv}) causes suppression of photon signatures above $\sim 0.1$~TeV. This implies that $\gamma$-rays produced in $\pi^0$-decays cannot serve as a signature of hadronic interactions. 
However, as shown in \citet{Rudolph:2022ppp}, internal $\gamma \gamma$-annihilation of hadronically produced VHE $\gamma$-rays may produce low-energy photon signatures distinct from purely leptonic models. This underlines the necessity to include a self-consistent treatment of particle cascades in dense sources such as GRBs when searching for photon signatures of hadronic interactions. \\
\textit{(2) Impact of intermediate pion and muon cooling.}
In contrast to photons, high-energy neutrinos are not subject to EBL absorption and may hence serve as direct signatures of hadronic interactions in a source. 
As can be inferred from Fig.~\ref{fig:GRB_example} both scenarios reproduce neutrino fluences below the reported IceCube limit.
We point out that for the IC-dominated scenario, the reduced synchrotron photon production efficiency implies a higher jet kinetic energy in order to reproduce the observed photon fluence. For the same baryonic loading, this translates into higher neutrino fluences. 
The shape of the neutrino spectra carries signatures of the cooling of intermediate particles in the source: for the SYN-dominated scenario of strong magnetic fields, pions and muons synchrotron cool prior to their decay. This introduces breaks in the neutrino spectra \citep[see, e.g.,][]{Lipari:2007su, Baerwald:2010fk, Baerwald:2011ee, Tamborra:2015qza, Bustamante:2020bxp}. On the other hand, for weaker magnetic fields (as in the case of the IC-dominated scenario), the neutrino spectrum is smoothly single-peaked.

\section{Efficiency}
\label{sec:efficiency}

We finish the discussion with a brief report on the software performance. The time necessary to perform an \AM simulation is divided into the calculation of the kernels and the evolution of the time-dependent solver by a given number of steps. We exclude from this discussion the overhead from binary loading time, module imports, and other auxiliary tasks. The \AM performance depends essentially on three factors, which we briefly discuss below.

\begin{figure}[H]
    \centering
    \includegraphics[width=0.7\linewidth]{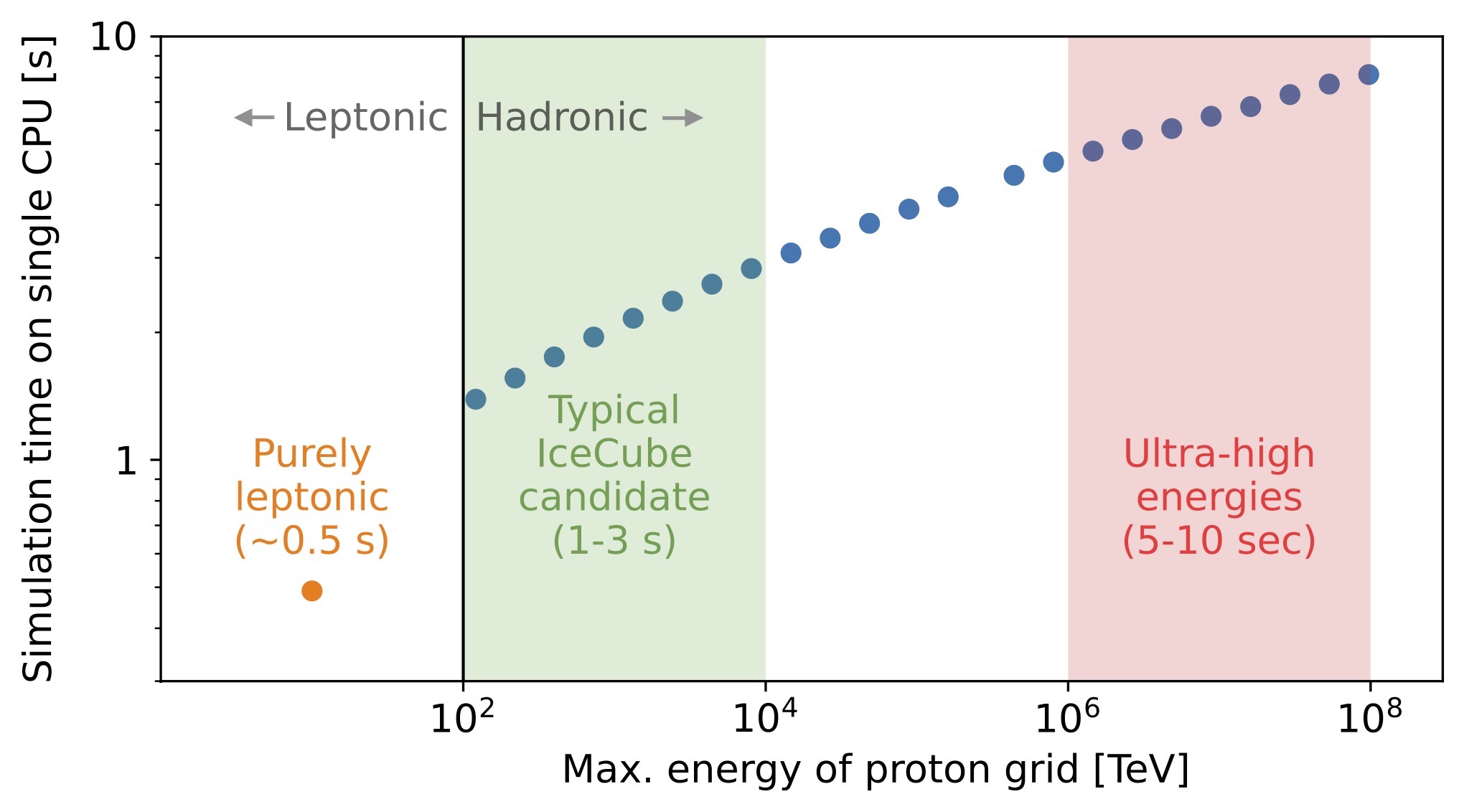}
    \caption{Typical \AM performance for leptonic (orange) and lepto-hadronic simulations (blue). The y-axis shows the total time necessary to run a full simulation on a single CPU using an Apple M2 chip. This includes the time necessary for initializing the kernels, creating and injecting the particle spectra, and subsequently running 30 steps of the time-dependent solver down to the steady state, but excludes any overhead from other accessory tasks.}
    \label{fig:blazar_performance}
\end{figure}

\textit{(1) What physical processes are considered.} This can be controlled by switches for individual phsyical processes and speed optimizing options, before initializing the kernels. 

As we can see in Fig.~\ref{fig:blazar_performance}, a leptonic simulation (i.e. including synchrotron emission and absorption, inverse Compton, and photon-photon pair production) will typically run in $\sim0.5~\rm{s}$, while the same simulation may take 10 times longer when proton interactions are included (Bethe-Heitler and photo-meson processes). The simulation time increases by an additional factor of $\sim3$ if muon and pion cooling are included, which may be important in extreme environments like GRBs, as discussed in Sec.~\ref{sec:applications_grb}.

For leptonic simulations, the simulation time is divided equally between kernel initialization and time solver evolution, while for hadronic simulations the calculation of the kernels dominates the overall execution time.

\textit{(2) Length of the photon, proton and electron energy grids}. This effect is shown in Fig.~\ref{fig:blazar_performance} for different proton energy grid lengths. We see that in the example highlighted in Sec.~\ref{sec:applications_blazar}, where the maximum proton energy grid was of about 100~PeV, the simulation takes an average of $\sim5$~s; for an ultra-high maximum proton grid energy of 100~EeV, that time increases to almost 10~s.

By default, the \AM energy grids are set up conservatively, in order to capture protons up to ultra-high energies. However, when efficiency is essential, such as for large parameter scans, the energy grids can be adjusted, within certain limits, as discussed in the documentation.

\textit{(3) Number of time steps.} In the example shown in Sec.~\ref{sec:applications_blazar}, the particles were considered to have speed-of-light escape. This means that the light-crossing time of the blob acts effectively as the dynamical timescale of the system, $R^\prime_\mathrm{b}=c\,t^\prime_\mathrm{dyn}$. At any given energy, the dominant cooling process for a given particle can be either faster than escape (i.e. lying below the black line in the middle panel of Fig.~\ref{fig:blazar_plots}), or at the longest escape itself. By running the system up to a time $t_\mathrm{sim}\gtrsim3t_\mathrm{dyn}$, we ensure that all relevant processes are captured, and that therefore the processes of injection, energy losses and escape have reached an equilibrium, leading to a steady-state solution where the particle densities no longer change in time (within a relative precision limited to a few per cent). It is important to note that while this condition has been tested in the case of this particular system, we cannot guarantee its generality regarding other simulated environments or regions of parameter space. For example, \citet{Petropoulou:2018sfc} and, more recently, \citet{Florou:2022okf}, have shown that under certain conditions leptohadronic systems can become supercritical, in which case the particle density evolution can exhibit a complex behavior in the time domain rather than exponentially approaching a steady state solution. Such regime can be achieved in environments with high particle densities, such as the prompt phase of GRBs, particularly for high values of the total energy in relativistic protons. This illustrates how the asymptotic behavior of the system in the time domain is reliant upon verification by the user through careful testing.

With this in mind, it becomes clear that the time step of the numerical solver, $\Delta t$, adopted by the user plays a crucial dual role: both in capturing the correct system behavior in the time domain, and regarding the efficiency of the framework, given that the run time scales linearly with the number of evolved steps. In the case of the blazar example, as well as the performance check of Fig.~\ref{fig:blazar_performance}, we utilize a time step $\Delta t=0.1t_\mathrm{dyn}$ and therefore evolved the solver by 30 time steps down to the steady state, $N=3t_\mathrm{dyn}/\Delta t$. In the other examples discussed previously, namely the TDE and GRB cases, interactions can be considerably faster, and therefore a smaller step size must generally be adopted in order to match the increasing stiffness of the differential equations. In general, an accuracy test is always advised before attempting to optimize the simulation by increasing the relative time step, in order to avoid stiffness-derived inaccuracies in the predicted SED.

\section{Summary and conclusion}
\label{sec:conclusions}

We have introduced \AM, an open-source numerical software for astroparticle simulations. It efficiently solves a set of time-dependent partial differential equations that self-consistently describe the spectral evolution of protons, neutrons, pions, muons, neutrinos, electrons/positrons and photons in a homogeneous and isotropic environment containing a magnetic field that can be updated at each time step.

It is possible to account for synchrotron and inverse Compton emission by electrons/positrons, protons, pions and muons, synchrotron self-absorption by electrons/positrons, photon-photon pair annihilation, Bethe-Heitler pair production, proton-proton and proton-photon pion production, as well as pion and muon decay. The software calculates both the energy losses of the parent particles and the emission of the respective products in a self-consistent and time-dependent fashion. For efficiency purposes, each of these processes can be individually switched on and off. The framework also simulates the physical escape of particles from the region of interest, as well as a possible time-dependent adiabatic expansion of that region. 

Particles of any of the above species can be injected into the simulation via a source term. In the case of electrons and protons, this typically represents pre-accelerated particles; in the case of photons, this feature can be used to emulate the presence of external radiation sources, such as a broad line region near an AGN. 

For capturing the time-dependent evolution, the software relies on a hybrid solver combining the speed, stability and accuracy of a tri-diagonal matrix method with analytical approximations for very fast cooling processes. 
For certain applications, the computational expense may further be reduced by adjusting the grid size (combined with interpolation) or omitting sub-dominant cooling processes, at a small loss in accuracy.

Since its initial application to the modeling of a neutrino source candidate \citep{Gao:2016uld}, \AM has subsequently been used to model the emission spectra of photons and neutrinos for several other blazars including the source TXS~0506+056~\citep{Gao:2018mnu,Rodrigues:2018tku,Rodrigues:2020fbu,Rodrigues:2023vbv,MAGIC:2023ybm}, as well as GRB prompt  \citep{Rudolph:2021cvn,Rudolph:2022ppp,Rudolph:2022dky, Rudolph:2023auv} and afterglow \citep{2024arXiv240313902K} emission, and more recently tidal-disruption events~\citep{Yuan_2023,2024arXiv240109320Y,2024arXiv240611513Y}. Additionally, the software has recently been extensively compared to other state-of-the-art frameworks~\citep{Cerruti:2021hah}, and good agreement has been found.

In the current release, the C++ code is distributed along with a Python3 interface with no loss of efficiency and providing access to all simulation features detailed in this work. 
The flexibility of \AM permits the user to build arbitrarily complex and diverse astrophysical models in a modular manner, see the various applications introduced before. Finally, \AM can be coupled to other physics codes, which would permit investigations of other phenomena such as interactions of cosmic-ray nuclei heavier than protons, combined source-propagation models, and physics beyond the Standard Model.

In the era of multi-messenger astrophysics, \AM provides an essential building block for modeling astrophysical high-energy sources. The framework allows to  utilize multi-messenger data to model the respective astrophysical environment and produce self-consistent predictions on cosmic rays, neutrinos, and multi-wavelength photons.

\section*{Acknowledgements}

We thank Maximilian Linhoff for his recent contribution to the \AM installation setup and Egor Podlesnyi for reporting an error in the blazar example affecting the external field conversion factor. M.K. is supported by the International Helmholtz-Weizmann Research School for Multimessenger Astronomy, largely funded through the Initiative and Networking Fund of the Helmholtz Association.
A.R. is supported by the Carlsberg Foundation (CF18-0183).
X.R. is supported by the Deutsche Forschungsgemeinschaft (DFG, German Research Foundation) through grant SFB 1258 ``Neutrinos and Dark Matter in Astro- and Particle Physics'' and by the Excellence Cluster ORIGINS, which is funded by the DFG under Germany's Excellence Strategy - EXC 2094 -- 390783311. 
This project has received funding from the
European Research Council (ERC), 
Grant Nos. 646623 (NEUCOS) and 949555 (MessMapp). 

\section*{Author contributions}

S.G. initiated the project and created the first version of \AM in C\texttt{++}, as in G16. He further implemented and tested the available options for performance optimization, the tracking of the different electron and photon components, as well as the updated Bethe-Heitler treatment and electron quantum synchrotron radiation. 
A.F. refactored the initial code to comply with an object-oriented coding style, enabling further modularization.
M.P. and W.W. participated in scientific discussions on particle interaction processes, targeting the software toward its applications.
M.K. has actively maintained and co-developed \AM. In particular, he has developed the current modularized structure of the C\texttt{++} library and user interface functions, developed the algorithm that automatically selects the appropriate solver, implemented the run time profiling and collaborated on the implementation of proton-proton interactions.
A.R. has actively maintained and co-developed \AM. In particular, she has conducted tests under various astrophysical contexts, collaborated on the implementation of proton-proton interactions, contributed conceptually to the modularization of the code and developed the \AM Python3 interface. She further collaborated on the creation of the comprehensive project documentation, accessible in the online repository.
X.R. has actively maintained and co-developed \AM. In particular, he has included the treatment of external photon fields and contributed conceptually to the code re-structuring. He further collaborated on the creation of the project documentation, and led the coordination efforts among the collaborators that led to the project release as open source.
C.Y. has actively maintained and co-developed \AM, wrote the documentation of the C\texttt{++} source code, and tested the code with C\texttt{++} scripts.
G.F.C. wrote the Docker interface for compiling and running \AM in a platform-independent environment, created the graphic user interface (GUI) that allows to run AM3 interactively, and designed the project logo.
The manuscript was written by A.R., M.K., X.R. and C.Y. with contributions from all remaining authors and utilizing some of the information provided by \citet{Gao:2016uld}.

\bibliography{references}{}
\bibliographystyle{aasjournal}

\appendix

\section{Treatment of particle interactions and other physical processes}
\label{app:physics_processes}
This section contains the mathematical treatment of the physical processes included in \AM, as well as the simplifications implemented to reduce the computation time (which can be switched on and off individually). For convenience, we here collect the descriptions for all processes, although the majority is already contained in G16. The below text thus largely follows G16, while extending/updating the prescriptions for processes where the code has been altered.

In the following, we use the standard symbols $p$ for protons, $n$ for neutrons, $\pi^\pm$ for charged pions, $\pi^0$ for neutral pions, $\mu^\pm_{R/L}$ for right ($R$) or left ($L$) handed muons, $\nu_{e/\mu}$/$\bar{\nu}_{e/\mu}$ for electron ($e$) or muon($\mu$) neutrinos/anti-neutrinos, $e^\mp$ for electrons/positrons and $\gamma$ for photons. 
Furthermore, particles of species $X$ with energy $E_X$ and mass $m_X$ will be described by their Lorentz factor $\gamma = E_X / (m_Xc^2)$, whereas for photons we use the dimensionless energy $\varepsilon \equiv E_\gamma / m_e c^2$ (with $m_e$ being the electron rest mass and $c$ the speed of light in vacuum). We recall our definition of $n = \dd^2 N / \dd E \dd V$, such that the total number of particles equals $N=\int dV\int dE n(E)$.

We furthermore generalize the Thomson cross-section $\sigma_T$ and critical magnetic field $B_\mathrm{crit}$ for charged particles of mass $m_X$ and charge number $Z = 1$ as:

\begin{equation}
    \sigma_\mathrm{T} = \frac{8 \pi}{3} \left(\frac{e^2}{m_X c^2}\right)^2,\hspace{5mm}
    B_\mathrm{crit}=\frac{m_X^2 c^3 }{e\hbar}, 
\end{equation}
where $\hbar$ is the reduced Planck constant and $e$ the electron electric charge.

\subsection{Synchrotron radiation}
Synchrotron radiation and cooling (labelled with subscript $syn$) due to movement in a magnetic field is implemented for all charged particles (electrons, positrons, protons, muons and pions), treating relativistic particles of $\gamma \gtrsim10$. \\
The emission rate of synchrotron photons with energy $\varepsilon$ created by species $X$ with Lorentz factor $\gamma_X$ is described through the integral

\begin{equation}
	Q_{\gamma, \mathrm{syn}}(\varepsilon)=\int \dd \gamma_X \: n_X(\gamma_X) R_\mathrm{syn}(\varepsilon, \gamma_X)  .
\end{equation}

The respective synchrotron kernels can be calculated following, e.g., \citet{Dermer:2009zz} (DM09):

\begin{equation}
	R_\mathrm{syn}(\varepsilon, \gamma_X) = \frac{3\sqrt{3}}{\pi} \frac{\sigma_\mathrm{T} c}{m_X c^2} \frac{B B_\mathrm{crit}}{8\pi} \left(\frac{m_X}{m_e}\right)^2 \hat{R}(z) = 2\sqrt{3} \frac{\alpha m_X c^2}{h} \left(\frac{m_X}{m_e}\right)^2\frac{B}{B_\mathrm{crit}}\, \hat{R}(z)
	\label{equ:Rsyn}
\end{equation}
where
\begin{align}
    \hat{R}(z) &= z^2 \left\{K_{4/3}(z)K_{1/3}(z)-\frac{3}{5}z\left[K_{4/3}^{2}(z)-K_{1/3}^{2}(z)\right]\right\}\\
    &\approx 
    \begin{cases}
        \frac{2^{-1/3}}{5} \Gamma(1/3)^2 z^{1/3}  \hspace{30mm} \mathrm{for} \; 2z < 10^{-2} \\
        \frac{1}{2} 10^{A_0 + A_1y + A_2 y^2 + A_3 y^3 + A_4 y^4 + A_5 y^5}  \hspace{5mm} \mathrm{for} \; 10^{-2} < 2z < 10 \; \mathrm{with} \; y= \log_{10}(2z)  \\
        \frac{\pi}{4} e^{-2z} \left( 1- \frac{99}{165}\frac{1}{2z}\right) \hspace{27mm} \mathrm{for} \;  2z > 10
    \end{cases}
\end{align}
and 
\begin{equation}
    z = \frac{\varepsilon m_e/m_X}{ 3\gamma_X^2 B/ B_\mathrm{crit}}
\end{equation}
and $K_{n}(z)$ is the modified Bessel function of the second kind. The coefficients $A_i$ are taken from table~1 of \cite{FinkeDermerBoettcher08_synapprox}. Note that $\sigma_\mathrm{T}$ and $B_\mathrm{crit}$ depend on the particle mass.

The cooling effect on charged particles is calculated as
\begin{equation}
	 - \dot{\gamma}_{X, \mathrm{syn}}=\dfrac{4}{3}\frac{\sigma_\mathrm{T} c}{m_Xc^2} \frac{B^2}{8\pi} \gamma_X^{2} = \frac{8\pi}{9} \frac{\alpha m_X c^2}{h} \left( \frac{B}{B_\mathrm{crit}} \right)^2 \gamma_X^2
\end{equation}

Synchrotron self-absorption is included only for electrons and positrons. The photon loss term is calculated as follows:
\begin{align}
	\alpha_{\gamma, \mathrm{ssa}}(\varepsilon) =\frac{\lambda_\mathrm{C}^{3}}{8\pi} \frac{1}{\varepsilon^2} \int \dd \gamma_e \:  R_\mathrm{syn}(\varepsilon, \gamma_e)  \gamma_{e}^{2}\partial_{\gamma_e}\left[\gamma_{e}^{-2}n_e(\gamma_{e})\right]
\end{align}
where $\lambda_\mathrm{C}=h/m_{e}c$ is the electron Compton wavelength. The term in the integral is expanded, $\gamma_{e}^{2}\partial_{\gamma_e}\left[\gamma_{e}^{-2}n_e(\gamma_{e})\right] = -3 \gamma_{e}^{-2} \left( \gamma_e n_e \right)  + \gamma_{e}^{-1} \partial_{\gamma_e} \left( \gamma_e n_e \right)$, and the derivative is calculated using a second-order centered difference scheme. 

Quantum synchrotron radiation introduces a suppression at the highest energies and is implemented following \citet{Brainerd_Petrosian:1987, Brainerd:1987} for ultra-relativistic electrons/positrons of $\gamma_{e^\pm} \gg 1$. The kernel is extracted from the power spectrum of a single particle, which is given by

\begin{equation}
 R_\mathrm{syn}(\varepsilon, \gamma_{e^\pm}) =  P_\mathrm{syn}(\varepsilon, \gamma_{e^\pm})= \sqrt{3} \alpha m_e c^2 F_B F_\xi\left(\frac{2 \varepsilon}{3 \gamma_{e^\pm}^2 F_B} \right)    
\end{equation}
 with 

\begin{equation}
F_\xi (x) = x \left[ \int_{z}^\infty K_{5/3}(x') dx' + zx \xi^2 K_{5/3}(z)) \right] \,  \hspace{1cm} \text{and} \hspace{1cm} F_B=\frac{B}{B_\mathrm{crit}}.
\end{equation}
Here we have used
\begin{equation}
    z = \frac{x}{1-x\xi},\hspace{5mm}
    \xi = \frac{3}{2 F_B \gamma_{e^\pm}},\hspace{5mm}
    x = \frac{\varepsilon}{ \gamma_{e^\pm} \xi} \, .
\end{equation}

\subsection{Inverse Compton scattering}
Similarly to synchrotron radiation, inverse Compton scattering (labelled by subscript IC) with photons is implemented for all charged particles. The incoming particle of species $X$ is represented by its Lorentz factors $\gamma_X$, while incoming and outgoing photons are represented by their dimensionless energies are $\varepsilon_i$ and $\varepsilon_o$ (normalised to the electron rest mass).
For each species $X$ the generation rate of photons can be obtained through

\begin{equation}
	Q_{\gamma, \mathrm{IC}}(\varepsilon_o)=\int \dd \varepsilon_i n_\gamma (\varepsilon_i) \: \int \dd \gamma_X \: n_X(\gamma_X) R_{X, \mathrm{IC}}(\gamma_X, \varepsilon_i, \varepsilon_o)  .
\end{equation}

For electron and positrons (with Lorentz factors $\gamma_{e^\pm}$) the description of the cross-section and the photon treatment are described in G16, considering only relativistic particles with $\gamma_{e^\pm}>10$. The differential cross-section (using $\sigma_{\mathrm{T}, e} = 8 \pi e^2 / 3 m_e^2 c^4$) can be much simplified under the head-on collision approximation (DM09)

\begin{equation}
	\frac{\dd^{2}\sigma_\mathrm{IC}}{\dd\varepsilon_{o}\dd\mu}\left(\gamma_{e^\pm},\varepsilon_{i},\varepsilon_{o},\mu\right)=
	\frac{3}{8}\frac{\sigma_{\mathrm{T}, e}}{\gamma_{e^\pm} x^{3}} \left[ yx^{2}+1+2x+\frac{1}{y}\left(x^2-2x-2\right)+\frac{1}{y^2} \right]
	H\left( y-\frac{1}{1+2x} \right) H\left(1-\frac{x}{2\gamma_{e^\pm}^2}-y\right)
\end{equation}
where $x=\gamma_{e^\pm} \varepsilon_{i}(1-\mu)$ is the photon energy in the particle rest frame, $y=1-\varepsilon_{o}/\gamma_{e^\pm}$ and $H(x)$ is the Heaviside step function. 
Performing the integration over the scattering angle $\mu$, the averaged inverse Compton kernel reads:
\begin{equation}
	\begin{aligned}
		R_{e^{\pm},\mathrm{IC}}\left(\gamma_{e^\pm}, \varepsilon_{i}, \varepsilon_{o}\right) 
			& =
				\dfrac{c}{2}\int \dd \mu (1-\mu) \: \dfrac{\dd^{2}\sigma_\mathrm{IC}}{\dd\varepsilon_{o}\dd\mu} \left(\gamma_{e^\pm},\varepsilon_{i},\mu \right) \\
		&	= 
            \begin{cases}
                c \sigma_{\mathrm{T}, e} \widetilde{R}(u,v)	& {\rm~for~} \dfrac{u}{2\gamma_{e^\pm}^{2}}<v<\dfrac{2u}{1+2u} {\rm ~and~} u<2\gamma_{e^\pm}^{2}-\frac{1}{2}\\
            	0							& {\rm otherwise}
            \end{cases} \\
	\end{aligned}
\end{equation}
\begin{equation*}
    {\rm~with~} \widetilde{R}(u,v) = 
            \dfrac{ 6(1-u)\dfrac{u}{v}\ln\left[\dfrac{u}{2v(1-u)}\right] - 3\left( \dfrac{u}{v}-2+2u \right) \left(\dfrac{u}{v}+1+\dfrac{u^{2}}{2}-u\right)}
					{4v\gamma_{e^\pm}(u-1)^{3}}
\end{equation*}

where $u=2\gamma_{e^\pm}\varepsilon_{i}$, $v=\varepsilon_{o}/\gamma_{e^\pm}$. We note that this is the same result as Eq.~2.49 in \cite{GouldBlumenthal1970}.

The total reaction rate, averaged over the angle $\mu$ and integrated in $\varepsilon_o$, is

\begin{equation}
	\begin{aligned}
		R_{e^{\pm}, \mathrm{IC}}(\gamma_{e^\pm},\varepsilon_{i})	&	 = \int \dd\varepsilon_{o} \: R_{e^{\pm}, \mathrm{IC}}\left(\gamma_{e^\pm}, \varepsilon_{i}, \varepsilon_{o}\right)  \\
													&	 = \dfrac{3c}{8u^{3}(1+2u)}\left[
														-2u(4+9u+u^{2})+(4+u)(1+2u)^{2}\ln(1+2u)+4u(1+2u)\mathrm{Li}_2(-2u) 
													\right]\; .
	\end{aligned}
\end{equation}

In this expression, ${\rm Li}(2,z)$ is the dilogarithm defined as
\begin{equation}
	\mathrm{Li}_2(z)=-\int_{0}^{z} \dd t \frac{\ln(1-t)}{t}
\end{equation}

The sink term for the photons due to upscattering by $e^\pm$ is calculated via the total interaction rate:
\begin{equation}
    \alpha_{\gamma, \mathrm{IC}} (\varepsilon_{i}) = \int \dd \gamma_{e^\pm} n_e(\gamma_{e^\pm}) R_{e^{\pm},\mathrm{IC}}\left(\gamma_{e^\pm}, \varepsilon_{i}\right) 
\end{equation}

For charged pions, muons and protons, inverse Compton scattering is implemented in a simplified version where the cross-section is a simple step function:
\begin{equation}
    \frac{\mathrm{d}\sigma_\mathrm{IC}}{\mathrm{d}\varepsilon_{o}} = \sigma_T \delta(\varepsilon_{o} - \gamma_{X,i}^2 \varepsilon_{i}) H \left( 1- \frac{1}{2}\gamma_{X, i} \varepsilon_{i} \frac{m_e}{m_X} \right) \, .
\end{equation}
Here we recall that $\sigma_T \propto m_X^{-4}$ is a function of the particle mass.

Finally, in contrast to G16, the loss is implemented through a continuous loss term for all particles. Following \citet{Jones:1965} for a particle of species $X$ it can be written as 

\begin{equation}
	- \dot{\gamma}_{X, \mathrm{IC}} = \frac{3}{16} \sigma_T c \int_0^\infty  \mathrm{d}\varepsilon_i \frac{n_{\gamma}(\varepsilon_i)}{\gamma_X^2 \varepsilon_i} F(\varepsilon_i, \gamma_X)
	\label{equ:ic_continuousloss} \, , 
\end{equation}
where 
\begin{equation}
    F(\varepsilon, \gamma_X) = \gamma_X \left[f_1 (\varepsilon \bar{\gamma}_X) - f_1 (\varepsilon/ \bar{\gamma_X})\right]
                                - \varepsilon \left[f_2 (\varepsilon \bar{\gamma}_X) - f_2 (\varepsilon/ \bar{\gamma}_X) \right]
\end{equation}

with 
\begin{align}
    \bar{\gamma}_X &= \gamma_X +(\gamma_X^2 -1)^{1/2} \\
    f_1(z) &= \qty(z+6+\frac{3}{z})\mathrm{\log}(1+2z) - \qty(\frac{22}{3}z^3 +24z^2 +18z +4)(1+2z)^{-2} -2 - 2 \mathrm{Li}_2(-2z) \\
    f_2(z) &= \qty(z + \frac{31}{6} + \frac{5}{z} + \frac{3}{2z^2})\mathrm{log}(1+2z) -\qty(\frac{22}{3}z^3 +28z^2 +\frac{103 z}{3} +17 + \frac{3}{z})(1+2z)^{-2} -2 - 2 \mathrm{Li}_2(-2z) \, , 
\end{align}
where in case of small and large $z$ the above expressions are simplified using a Taylor expansion around $z = 0$.

\subsubsection*{Performance optimization}
Several optimization options are available to reduce the computational time:
\begin{itemize}
    \item Reduction of the grid of out-going photons by factor 2 by using only every second energy bin. The values in between are interpolated.
    \item Usage of only every second bin for in-going photons.
    \item Adjustment of the minimum energy of in-going photons considered for the interaction. The pre-set value is an energy of $\sim 7.8 \times 10^{-3}$~eV. 
    \item Adjustment of the maximum energy of in-going photons considered for the interaction. The pre-set value is an energy of $m_e c^2 = 511$~keV.
\end{itemize}

\subsection{Photon-photon pair annihilation}
The annihilation of two photons into an electron-positron pair (labelled by subscript $pair$) is described by the reaction $\gamma +\gamma \to e^+ + e^-$. We denote with the indices 1 and 2 the dimensionless energies $\varepsilon_1$, $\varepsilon_2$ of the incoming photons and the electron ($e^-$) and positron ($e^+$) have Lorentz factors $\gamma_{e^\pm}$.
The treatment of the process is the same as in G16 \citep[which, in turn, follows][]{Vurm:2008ue}).
Energy conservation implies $\varepsilon_{1}+\varepsilon_{2} = \gamma_{e^-}+\gamma_{e^+}$. We start by describing the source terms for electrons/positrons, which are symmetric:

\begin{equation}
	Q_{e^\pm, \mathrm{pair}}(\gamma_{e^\pm})=\int_{\varepsilon_1^{*}}^{\infty} \dd \varepsilon_1 \: n_\gamma(\varepsilon_1) \int_{\varepsilon_2^{*}}^{\infty} \dd \varepsilon_2 \: n_\gamma(\varepsilon_2) \:  R_{e^\pm, \mathrm{pair}} (\varepsilon_1,\varepsilon_2,\gamma_{e^\pm})
	\label{equ:Qpp}
\end{equation}
where, 
\begin{equation}
	R_{e^\pm, \mathrm{pair}} (\varepsilon_1,\varepsilon_2,\gamma_{e^\pm}) 
    =- \sigma_{\mathrm{T},e c}\frac{\left[S(\gamma_{e^\pm},\varepsilon_2,\varepsilon_1,w_{U})-S(\gamma_{e^\pm},\varepsilon_2,\varepsilon_1,w_{L})\right]}{4\varepsilon_1^{2}\varepsilon_2^{2}}
\end{equation}
in which,
\begin{equation}
	S(\gamma_{e^\pm},\varepsilon_2,\varepsilon_1,w)=-\left[(\varepsilon_1+\varepsilon_2)^{2}-4w^{2}\right]^{1/2} + T(\gamma_{e^\pm},\varepsilon_2,\varepsilon_1,w) + T(\gamma_{e^\pm},\varepsilon_1,\varepsilon_2,w)
\end{equation}
where,
\begin{equation}
	\begin{aligned}
	T(x,y,z,w) =& \frac{w^{3}}{(y z)^{3/2}} \frac{(y z - 1)}{h} \left[A_{0}(h)-(1+h)^{1/2}\right]-\frac{1}{w} \left(\frac{1+h}{yz} \right)^{1/2} \\ 
				&+ \frac{w}{2(y z)^{3/2}} \left[(1+h)^{-1/2}(y^{2}+y z +x z - x y -2w^{2})-4 y z A_{0}(h) \right]\\
	\end{aligned}
\end{equation}
where,
\begin{equation}
	A_{0}(h)
		\begin{cases}
			= h^{-1/2}\ln\left[h^{1/2}+(1+h)^{1/2}\right] & {\rm ~ for ~} h>0 \\
			=- h^{-1/2} \arcsin\sqrt{-h} & {\rm ~ for ~} h<0 \\
			\approx 1-h/6+3h^{2}/40 & {\rm ~ for ~} h\approx0 \\
		\end{cases}
\end{equation}
and finally, with $h=\left[(x-y)^{2}-1\right]w^{2}/y z$ and the integration boundaries $w_{L}=w_{-}$, $w_{U}=\min\left[\sqrt{\varepsilon_1\varepsilon_2},w_{+}\right]$, in which, 
\begin{equation}
	w_{\pm}= \frac{1}{2}\left[ \gamma_{e^\pm}\gamma_{e^\mp} + 1 \pm \sqrt{(\gamma_{e^\pm}^{2}+1) (\gamma_{e^\mp}^{2}+1)} \right] \ \hspace{1cm} {\rm~with~} \gamma_{e^\mp} = \varepsilon_{1}+\varepsilon_{2} - \gamma_{e^\pm} .
\end{equation}
The lower-limits of the integration in \equ{Qpp} are
\begin{equation}
	\begin{aligned}
		\varepsilon_{1}^{*} & =\dfrac{1}{2}\gamma_{e^\pm}(1-\beta_{e^\pm}) \\ 
		\varepsilon_{2}^{*} & = 
            \begin{cases}
                \varepsilon_{2}/\{[2\varepsilon_{2}-\gamma_{e^\pm}(1+\beta_{e^\pm})]\gamma_{e^\pm}(1+\beta_{e^\pm})\} & {\rm~for~} x>x_{2}\\
                \varepsilon_{2}/\{[2\varepsilon_{2}-\gamma_{e^\pm}(1-\beta_{e^\pm})]\gamma_{e^\pm}(1-\beta_{e^\pm})\} & {\rm~for~} x<x_{1}\\
                \gamma_{e^\pm}-\varepsilon_{2}+1 & {\rm~for~} x_{1}\le x \le x_{2}\\
			\end{cases}
	\end{aligned}
\end{equation}
with $x_{1}=[1+\gamma_{e^\pm}(1-\beta_{e^\pm})]/2$, $x_{2}=[1+\gamma_{e^\pm}(1+\beta_{e^\pm})]/2$ and $\beta_{e^\pm} = \left[ 1 - \gamma_{e^\pm}^{-2} \right]^{1/2}$.

The disappearance rate for photons with $\varepsilon_{2}$ due to pair production with target photons with $\varepsilon_{1}$ is
\begin{equation}
	\alpha_{\gamma, \mathrm{pair}}(\varepsilon_{2})
 = \int \dd\varepsilon_{1} \: n_\gamma(\varepsilon_{1}) \int \dd\gamma_{e^\pm} \: R_\mathrm{pair}(\varepsilon_{2},\varepsilon_{1}, \gamma_{e^\pm})
 = \int \dd\varepsilon_{1} \: n_\gamma(\varepsilon_{1}) \: R_{\gamma, \mathrm{pair}}(\varepsilon_{2},\varepsilon_{1})
\end{equation}
where $R_{\gamma, \mathrm{pair}}(\varepsilon_{2},\varepsilon_{1})$ is given by (see also DM09)
\begin{equation}
	R_{\gamma, \mathrm{pair}}(\varepsilon_{2},\varepsilon_{1})=\dfrac{3}{8} \frac{\sigma_{\mathrm{T}, e} c}{\varepsilon_{2}^{2}\varepsilon_{1}^{2}} \bar{\varphi}(u,v)
\end{equation}
with 
\begin{equation}
	\bar{\varphi}(u,v)=\left[2v+(1+v)^{-1}\right]\ln u - \ln^{2}u - \frac{2(2v+1)v^{1/2}}{(v+1)^{1/2}} + 4\ln u \ln(1+u) + \frac{\pi^{2}}{3}-4\left[{\rm Li}_{2}(-1)-{\rm Li}_{2}(-u)\right]
	\label{equ:phibaruv}
\end{equation}
and 
\begin{equation}
	v=\varepsilon_{2}\varepsilon_{1}-1, \hspace{5mm} u=\dfrac{\sqrt{v+1}+\sqrt{v}}{\sqrt{v+1}-\sqrt{v}} \, .
\end{equation}
The asymptotic form of \equ{phibaruv} is 
\begin{equation}
 	\bar{\varphi}(u,v)\approx\begin{cases}
 		2v\ln(4v)-4v+\ln^{2}(4v) & {\rm ~for~} v>>1 \\
 		\dfrac{4}{3}v^{3/2} & {\rm ~for~} v<<1
		\end{cases}\ .
\end{equation}
The rates for the second photon participating in the reaction are obtained by switching $\varepsilon_1$ and $\varepsilon_2$.

\subsubsection*{Performance optimization}
 Computational time may be reduced by using an 14-bin approximation for the integration kernel. 
 For this, we assume that the energy budget is dominated by the high-energy photon participating in the interaction (e.g., $\varepsilon_2 \gg \varepsilon_1$).
 The created pairs are re-distributed over 14 bins (consistent with our energy resolution of $\Delta \ln E = 0.1$, which implies a distance of 7 bins between $\varepsilon_2$ and $\varepsilon_2/2$):
    \begin{align}
        Q_{e^\pm, \mathrm{pair}, 14 \mathrm{bin}}(\gamma_{e^\pm}) = \int_{1}^{\infty} \mathrm{d} \varepsilon_2 \: n_\gamma(\varepsilon_2) \: \left[ \sum_{i=0}^{i=6} \delta\left(\gamma_{e^\pm} - \gamma_{k,+} \right)  +  \sum_{i=0}^{i=7}  \delta\left(\gamma_{e^\pm} - \gamma_{k,-} \right) \right ] \: \frac{\alpha_{\gamma, \mathrm{pair}}(\varepsilon_2)}{16} \, ,
    \end{align}
where $\gamma_{k,+} =  \varepsilon_2 \exp(k \Delta x) / 2 $ and $\gamma_{k,-} = \varepsilon_2 / 2 \left(2 - \exp(k \Delta x)\right)$ (with the logarithmic energy grid spacing $\Delta x = \Delta \ln E = 0.1$). Note that the above formula implies that \textit{(1)} the injection at energy $\varepsilon_2/2$ is enhanced by a factor of 2 and \textit{(2)} reinjection at the lowest grid point is neglected.

\subsection{Bethe-Heitler pair production}

The production of lepton pairs due to Bethe-Heitler pair production in the field of protons ($p + \gamma \rightarrow p + e^+ + e^- $, labelled by subscript $BH$) is implemented following \citet{Kelner:2008ke}. For the interaction between a proton of Lorentz factor $\gamma_p$ and a photon of dimensionless energy $\varepsilon$ the differential electron generation rate $Q_{e^\pm}(\gamma_e)$ is given by

\begin{equation}
    Q_{e^\pm, \mathrm{BH}}(\gamma_e) = \frac{3 }{16 \pi } \alpha \sigma_T c \int \dd \varepsilon \: n_\gamma(\varepsilon) \int \dd \gamma_p \:  n_p(\gamma_p)  \: R_\mathrm{BH}(\gamma_p, \varepsilon, \gamma_e) \, .
\end{equation}

We calculate the integration kernel as in Eq.~62 of \citet{Kelner:2008ke}: 

\begin{equation}
     R_\mathrm{BH}(\gamma_p, \varepsilon, \gamma_e) = \frac{\gamma_e}{2 \gamma_p^3 \varepsilon^2} \int_{(\gamma_p + \gamma_e)^2 / (2 \gamma_p \gamma_e)}^{2 \gamma_p \varepsilon} \mathrm{d}\omega \: \omega \: \int_{\frac{\gamma_p^2 +\gamma_e^2}{2 \gamma_p \gamma_e}}^{\omega -1 } \frac{\mathrm{d} E_{-}}{p_{-}} \sigma_W(\omega, E_{-}, \xi) \, 
\end{equation}
where $E_{-}$ is the energy of the electron in the proton restframe, $p_{-} = \sqrt{E_{-}^2 -1}$ and $\xi = (\gamma_p E_{-} - \gamma_e)/\gamma_p p_{-}$.
The cross-section $\sigma_W \equiv \mathrm{d}\sigma^2/\mathrm{d}E_{-} \mathrm{d} \cos\theta_{-}$ is implemented following Eq.~10 in \citet{1970:Blumenthal}. These expressions were derived in the Born limit of $\gamma_p \varepsilon m_e \ll m_p$, which in our case is implemented as a conservative upper limit of $\omega < 600$. 

The energy loss rate is defined from Eq.~9.35 of DM09:
\begin{equation}
    - \dot{\gamma}_{p, \mathrm{BH}} (\gamma_p, \varepsilon) = \frac{3}{8\pi} \alpha \sigma_{\mathrm{T},e} c \frac{m_e}{m_p} \int_2^{\infty} \dd\varepsilon \: n_\gamma \left(\frac{\varepsilon}{2\gamma_p} \right) \:\frac{\phi(\varepsilon)}{\varepsilon^2} \, ,
\end{equation}
where the function $\phi(\varepsilon)$ is fitted as 

\begin{align}
    \phi(\varepsilon) = \begin{cases}
        \dfrac{\pi}{12} \dfrac{(\varepsilon -2 )^4}{1 + \sum_{i=1}^4c_i(\varepsilon-2)^i} \, &, \mathrm{for \, }2\leq \varepsilon < 25 \\
        \dfrac{\varepsilon \sum_{i=1}^4 d_i(\mathrm{ln}(\varepsilon)^i}{1-\sum_{i=1}^3f_i\varepsilon^{-1}} \, &, \mathrm{for \, } 25 \leq \varepsilon 
    \end{cases} \, .
\end{align}
The coefficients are given as 
$(c_1, c_2, c_3, c_4) = (0.8048, 0.1459, 1.137 \times 10^{-3}, -3.879 \times 10^{-6})$, 
$(d_1, d_2, d_3, d_4) = (-86.07, 50.96, -14.45, 8/3)$ and
$(f_1, f_2, f_3) = (2.91, 78.35, 1837)$ .

\subsubsection*{Performance optimization}
Computational time may be reduced by switching on one or more of the below optimization options:
\begin{itemize}
    \item Exact calculation of only every second bin of the out-going electrons, while interpolating the in-between values.
    \item Adjustment of the minimum energy of the in-going protons considered for the interaction. The default value for this switch is $1$~TeV. 
    \item Adjustment of the maximum energy of the in-going photons considered for the interaction: the default value for this switch is the electron rest mass, so $511$~keV. 
    \item As it is often sub-leading, the computation time may further be reduced by omitting the cooling term for protons in the calculations.
\end{itemize}

\subsection{Photo-pion process}
The photo-pion process (labelled by the subscript $p\gamma$) is implemented as described in G16, following the simplified treatment \textit{sim A} of \citet{Hummer:2010vx} (H10). \AM includes the interactions between protons and photons as well as between neutrons and photons, but here we limit ourselves to giving the treatment for protons. The neutron terms are conceptually similar, using different values for the parameters of the same parameterization.
We denote the incoming proton Lorentz factor $\gamma_p$, the incoming photon's dimensionless energy $\varepsilon$ and $\gamma_X$ represents the Lorentz factor of the outgoing particle that can be a proton, neutron or pion.

The generation rate of particle $X$ can be expressed as
\begin{equation} 
	Q_{X, \mathrm{p\gamma}}(\gamma_{X})=\int \dd \varepsilon n_\gamma(\varepsilon) \: \int \dd \gamma_p n_p(\gamma_p) \:  R_\mathrm{p\gamma} (\gamma_p, \varepsilon, \gamma_X)  .
\end{equation}

Based on H10, the integration kernel is parameterized as a sum over interaction channels $i$. For each kernel $R_\mathrm{p\gamma, i}$, the angle-averaged, differential cross-sections factorized is into a $\delta$-function, the multiplicity $M_i$ of the secondary particle $X$ and a function $f_i(y)$ depending only on $y=\gamma_{p}\varepsilon$ and containing the angular averaging:
\begin{equation} \label{eq:pgamma_delta}
    R_\mathrm{p\gamma} (\gamma_p, \varepsilon, \gamma_X) = \sum_{i} R_\mathrm{p\gamma, i} (\gamma_p, \varepsilon, \gamma_X) = c\gamma_p^{-1}\sum_{i}\delta \left( \frac{\gamma_X}{\gamma_p}-\chi_{i}\right)M_{i}f_{i}(y) ,
\end{equation}
where $\chi_{i}$ is the inelasticity of the collision. The functions $f_i(y)$ are stored in a look-up table during initialization and defined as 
\begin{equation}
	f_i(y)=\dfrac{1}{2y^{2}}\int_{\epsilon_{th, i}}^{2y} d\epsilon_{r}\epsilon_{r}\sigma_i(\epsilon_{r}) . 
\end{equation}
We use the notation $Q\equiv \dd n_{X}/\dd\gamma_{X} \dd t$ and $n_\gamma\equiv \dd^{2}N_\gamma/\dd V \dd \varepsilon$, so that the expressions of $f_{i}(y)$ and coefficients can be directly applied as given by Eq.~30, 33-35 \& 40 and table 3, 5 \& 6 of HU10. Consequently, $i$ falls into the category of resonance, direct and multi-pion production, respectively.

Evaluating the $\delta-$function in Eq.~\ref{eq:pgamma_delta} on the integration over the proton Lorentz factor, we obtain:
\begin{equation}
	Q_{X, \mathrm{p\gamma}}(\gamma_X)=c \: n_{p}\left(\chi_{i}^{-1}\gamma_X \right) \: \gamma_X^{-1}m_{p}\sum_{i}M_{i}\int \dd y \: n_\gamma \left(\gamma_X^{-1}\chi_{i}m_{p}y\right) f_{i}(y).
	\label{equ:QxEx}
\end{equation}

The disappearance rate due to participation in $p\gamma$ process for protons is 
\begin{equation}
	\begin{aligned}
        \alpha_{p, \mathrm{p\gamma}}(\gamma_p)	
        = & \int \dd\gamma_X \int \dd\varepsilon  R_\mathrm{p\gamma} (\gamma_p, \varepsilon, \gamma_X) \\
        = & c \sum_{i}M_{i}\int \dd\varepsilon n_\gamma(\varepsilon)f_{i}(\gamma_{p}\varepsilon)
	\end{aligned} \, .
\end{equation}
For photons the disappearance rate reads
\begin{align}
	\alpha_{\gamma, \mathrm{p\gamma}}(\varepsilon) = & \int \dd\gamma_X n_X(\gamma_X) \int \dd \gamma_p n_{p}(\gamma_p)  R_\mathrm{p\gamma} (\gamma_p, \varepsilon, \gamma_X) \\
 = &
 c\sum_{i}\chi_{i}^{-1}M_{i}\int d\gamma_Xf_{i}(\chi_{i}^{-1}\gamma_X\varepsilon m_{p}^{-1})n_{p}(\chi_{i}^{-1}\gamma_X).
\end{align}

\subsubsection*{Performance optimization}
Computational time may be reduced by switching on one or more of the below optimization options:
\begin{itemize}
    \item Usage of only every second bin for the in-coming photons.
    \item Adjustment of the maximum photon energy considered in the reaction. The pre-set value is the electron rest mass, so $511$~keV.
    \item As it is often not the dominant cooling term for photons, computation time may further be reduced by omitting the photon cooling term in the calculations.
\end{itemize}

\subsection{Inelastic Proton-Proton collisions}

Inelastic proton-proton interactions (labelled as $pp$) between a target distribution of thermal (i.e. non-relativistic) protons and relativistic protons are implemented mainly following \citet{Kelner:2006tc}. 
The source term for pions is thus retrieved by performing the integration over the proton distribution:
\begin{equation}
    Q_{\pi, \mathrm{pp}} (\gamma_\pi) = n_{p, \mathrm{th}} \int \mathrm{d}\gamma_p n_p(\gamma_p) R_\mathrm{pp}(\gamma_p,\gamma_\pi)  \, , 
\end{equation}
where $n_{p, \mathrm{th}}$ is the density of thermal target protons and $R_\mathrm{pp}(\gamma_p,\gamma_\pi)$ is given as 

\begin{equation}
    R_\mathrm{pp}(\gamma_p,\gamma_\pi) = c \sigma_\mathrm{pp, \mathrm{inel}}(\gamma_p) \frac{1}{ \gamma_p } F_{\pi} \qty(\gamma_p, \frac{\gamma_\pi m_\pi}{\gamma_p m_p}) .
\end{equation}
The redistribution function $F_\pi (\gamma_p, x)$, where $x=E_\pi/E_p $, can either be calculated either using the \textsc{QGSJET} or \textsc{SYBILL} parametrization. 
For \textsc{QGSJET} it reads 

\begin{equation}
    F_{\pi, \mathrm{QGSJET}} (\gamma_p, x) = 4 \alpha B_\pi x^{\alpha -1} \left( \frac{1 - x^\alpha}{(1+r x^\alpha)^3} \right)^4 \times 
    \left(\frac{1}{1-x^\alpha} + \frac{3 r}{1 + rx^\alpha}\right) \times \left(1- \frac{m_\pi}{x \gamma_p m_p} \right)^{1/2}
\end{equation}
 with the parameters $B_\pi = 5.58 + 0.78 L + 0.1 L^2$,  $r = 3.1 / B_\pi^{3/2}$, $\alpha = 0.89 / \qty[B_\pi^{1/2} \qty(1 - \mathrm{exp}(-0.33 B_\pi))]$ as a function of normalized proton energy $L = \mathrm{ln}(\gamma_p m_p c^2 / 1 \mathrm{TeV})$. \\
 Instead, the \textsc{SYBILL} parametrization is

\begin{equation}
    F_{\pi, \mathrm{SYBILL}} (\gamma_p, x) = 4 \alpha B_\pi x^{\alpha -1} \left( \frac{1 - x^\alpha}{1+r x^\alpha(1 - x^\alpha)} \right)^4 \times 
    \left(\frac{1}{1-x^\alpha} + \frac{r(1 - 2x^\alpha)}{1 + rx^\alpha(1 - x^\alpha)}\right) \times \left(1- \frac{m_\pi}{x \gamma_p m_p} \right)^{1/2}
\end{equation}
 with the parameters $B_\pi = a + 0.25 $,  $r = 2.6 / a^{1/2}$, $\alpha = 0.98/ a^{1/2}$ as a function of $a = 3.67 + 0.83 L + 0.075 L^2$ and normalized proton energy $L = \mathrm{ln}(\gamma_p m_p c^2 / 1 \mathrm{TeV})$. \\

The updated cross-section presented in \citet{Kafexhiu:2014cua} reads

\begin{equation}
    \sigma_\mathrm{pp, \mathrm{inel}} (\gamma_p) = \left[ 30.7  - 0.96 \mathrm{log}\qty(\frac{T_p}{E_\mathrm{th}}) +0.18 \mathrm{log}^2\qty(\frac{T_p}{E_ \mathrm{th}}) \right ] \times \left[1 - \qty(\frac{T_p}{E_\mathrm{th}})^{1.9} \right]^3 \, \mathrm{mb}.
    \label{equ:sigma_pp}
\end{equation}
Here $E_\mathrm{th} = 2 m_\pi + m_\pi^2 / 2 m_p$ is the threshold energy depending on the pion mass and $T_p = (\gamma_p -1) m_p c^2$ is the proton kinetic energy.

The energy-dependent inelasticity $\kappa$ for charged and neutral pion production, needed for the proton energy losses and the delta approximation, is calculated as

\begin{align}
    \kappa_{\pi^{\pm}} (\gamma_p) &= \int \mathrm{d} \gamma_\pi F_{\pi} \qty(\gamma_p, \frac{\gamma_\pi m_{\pi^\pm}}{\gamma_p m_p}) \frac{\gamma_\pi m_{\pi^\pm}^2}{\gamma_p (\gamma_p -1) m_p^2},\\
    \kappa_{\pi^{0}} (\gamma_p) &= \int \mathrm{d} \gamma_\pi F_{\pi} \qty(\gamma_p, \frac{\gamma_\pi m_{\pi^0}}{\gamma_p m_p}) \frac{\gamma_\pi m_{\pi^0}^2}{\gamma_p (\gamma_p -1) m_p^2}.
\end{align}

Below a user-defined transition proton energy $E_{\mathrm{trans}, \delta}$ (that is pre-set to 100~GeV), we use the $\delta$-function approximation (remember $m_\pi \gamma_\pi = x \gamma_p m_p$):
\begin{equation}
    F_\pi (\gamma_p, x) = \frac{\gamma_p}{\kappa_\delta} \delta \qty( \gamma_p -1 - x \gamma_p \frac{1}{\kappa_\delta} \frac{m_{\pi}}{m_p}).
\end{equation}
For the $\delta$-function description this results in the injection of pions, calculated following Eq.~77 of \citet{Kelner:2006tc}:  
\begin{equation}
    Q_{\pi, \mathrm{pp}} (\gamma_\pi) =  n_{p, \mathrm{th}} c \sigma_\mathrm{pp, \mathrm{inel}}(E_{p, \mathrm{parent}}) \frac{n_p(E_{p, \mathrm{parent}})}{\kappa_\delta }\, , 
\end{equation}
where $E_{p, \mathrm{parent}} = m_p c^2 + \gamma_\pi m_\pi c^2/\kappa_{\delta}$ is the parent proton energy for a given pion energy in the $\delta$-function approximation. 
For a smooth transition, the inelasticity $\kappa_{\delta}$ is set to the inelasitcity at the transition energy, e.g., $\kappa_{\delta} \equiv \kappa_{\pi^0} (E_{\mathrm{trans}, \delta})$.

The cooling rate of protons is defined as 
\begin{equation}
    \dot{\gamma}_{p, \mathrm{pp}} (\gamma_p)=\qty[2 \kappa_{\pi^{\pm}} \sigma_\mathrm{pp, \pi^\pm, \mathrm{inel}} (\gamma_p) + \kappa_{\pi^{0}} \sigma_\mathrm{pp, \pi^0, \mathrm{inel}} (\gamma_p)] n_{p, \mathrm{th}} c \gamma_p \, .
\end{equation}
To obtain $\sigma_\mathrm{pp, \mathrm{inel}}$ for $\pi ^{\pm}$ ($\sigma_\mathrm{pp, \pi^\pm, \mathrm{inel}}$) and $\pi^0$ ($\sigma_\mathrm{pp, \pi^0, \mathrm{inel}}$), the respective pion masses are plugged in Eq.\ref{equ:sigma_pp}.

For the generation of $\pi_0$-decay photons, it is alternatively possible to follow the parametrization of \citet{Kafexhiu:2014cua}, implemented as a direct photon source term:
\begin{equation}
    Q_{\gamma, \mathrm{pp}} (\varepsilon) = n_{p, \mathrm{th}} \int \mathrm{d}\gamma_p n_p(\gamma_p) \: c\: \frac{\dd \sigma_\mathrm{pp\to \pi^0 \to \gamma \gamma} }{\dd \varepsilon}\qty(\gamma_p, \varepsilon).
\end{equation}

Here we used the implementation\footnote{\url{https://sourceforge.net/projects/lippgam/files/}}, allowing for the different parameterizations SYBILL, QGSJET, Pythia8 and Geant4. It follows the structure of a normalization factor times the redistribution function:
\begin{equation}
    \frac{\dd \sigma_\mathrm{pp\to \pi^0 \to \gamma \gamma} }{\dd \varepsilon}\qty(\gamma_p, \varepsilon) =    A_{\mathrm{max}} (T_p ) \times F(T_p , E_\gamma) \; .
\end{equation}
The functions are defined in terms of proton kinetic energy $T_p=(\gamma_p-1) m_pc^2$ and photon energy $E_\gamma = \varepsilon m_ec^2$. Given their rather complex form by combining different fitted parameterizations at different energy regimes, we refer the reader to \citet{Kafexhiu:2014cua} for the complete, detailed description.

\subsection{Adiabatic expansion}

Adiabatic expansion (labelled as $exp$) results in particle cooling and plasma dilution. If the expansion timescale is $t_\mathrm{exp}$ (defined from the volume expansion rate $\dot{V}$ as $t_\mathrm{exp} = V/\dot{V}$), the respective adiabatic cooling and volume dilution sink terms for a particle of Lorentz factor $\gamma_X$ (dimensionless energy $\varepsilon =E_\gamma / m_e c^2$ for photons and neutrinos) are defined as 

\begin{align}
    \alpha_{X,\mathrm{exp}} (\gamma_X) &= \alpha_{\gamma,\mathrm{exp}} (\varepsilon) = \frac{1}{t_\mathrm{exp}} \\
    \dot{\gamma}_{X,\mathrm{ad}} (\gamma_X) &= \frac{\gamma_X}{3 t_\mathrm{exp}} \, .
\end{align}

Note that due to their weak coupling with the collisionless plasma, neutral particles (neutrons, neutrinos and photons) are implemented to experience only a sink term $\alpha_{\gamma,\mathrm{exp}}$ and no cooling term $\dot{\gamma}_{X,\mathrm{adi}}$. 

The definition of the sink term can be motivated by total particle number conservation [$\mathrm{d} (nV)/ \mathrm{d} t \overset{!}{=} 0$]:
\begin{equation}
\frac{\mathrm{d}n}{\mathrm{d}t} = - \frac{n}{V}\frac{\mathrm{d}V}{\mathrm{d}t} = - \frac{1}{t_\mathrm{exp}} n \, .
\end{equation}
The cooling term may be retrieved recalling the definition of adiabatic expansion (
$\gamma_X V^{\hat{\gamma}- 1} \overset{!}{=} \mathrm{const.}$, where $\hat{\gamma} = 4/3$ is the adiabatic index). 
Note that \AM only deals with densities, such that an expanding volume is completely accounted for by continuously updating the expansion time scale. The volume parameter in the injection template is only used once to calculate the injected density more conveniently, but does not affect any other part of the code.

\subsection{Pion and muon decay}
Pion and muon decays (labelled as $dec$) are implemented through the channels
\begin{align}
		\pi^{+}(\pi^{-}) & \rightarrow \mu^{+}(\mu^{-})+\nu_{\mu}(\bar{\nu}_\mu) \\
		\mu^{+}(\mu^{-}) & \rightarrow e^{+}(e^{-})+\nu_{e}(\bar{\nu}_{e})+\bar{\nu}_{\mu}(\nu_{\mu})
	 \, .
\end{align}

The implementation follows \citep{Lipari:2007su}. For a particle $X$ with energy $E_X$ decaying into a particle $Y$ of energy $E_Y$ the source term is described as, 
\begin{equation}
    Q_{Y,\mathrm{dec}} (E_Y) = \int \dd E_X n_X(E_X) \: R_{X \to Y, \mathrm{dec}}(E_X, E_Y)
\end{equation}
with the scaling relation
\begin{equation}
    R_{X \to Y,\mathrm{dec}}(E_X, E_Y) = \frac{1}{E_X t_{X,\rm dec}} \frac{f_Y(\frac{E_Y}{E_X})}{E_X} \, 
    \label{equ:R_decay}.
\end{equation}
Here, $t_{X,\rm dec}$ is the decay timescale for the parent particle (in its rest frame). The probability density function $f_Y(x)$ as a function of scaling variable $x=E_{Y}/E_{X}$ for muons is given as 
\begin{equation}
	\begin{aligned}
		f_{\mbox{$\mu$}_{\mbox{\scriptsize{$R$}}}^{\mbox{\small {$+$} } } }(x) =
		f_{\mbox{$\mu$}_{\mbox{\scriptsize{$L$}}}^{\mbox{\small {$-$} } } }(x) =
		\dfrac{r_{\mu\pi}^{2}(1-x)}{(1-r_{\mu\pi}^{2})^{2}x}H(x-r_{\mu\pi}^{2}), \\
		f_{\mbox{$\mu$}_{\mbox{\scriptsize{$L$}}}^{\mbox{\small {$+$} } } }(x) =
		f_{\mbox{$\mu$}_{\mbox{\scriptsize{$R$}}}^{\mbox{\small {$-$} } } }(x) =
		\dfrac{x-r_{\mu\pi}^{2}}{(1-r_{\mu\pi}^{2})^{2}x}H(x-r_{\mu\pi}^{2})  ,
	\end{aligned}
\end{equation}
with $r_{\mu\pi} = m_\mu / m_\pi $. 
Note that internally, \AM distinguishes between left- and right-handed (anti-)muons since the decays are helicity-dependent. 

Due to energy conservation, the corresponding neutrino directly from pion decay has a distribution function of
\begin{equation}
	f_{\nu}(1-x)=f_{\mu}(x)
\end{equation}
For $\pi^{0}\rightarrow\gamma+\gamma$, the distribution function for a photon is (see, e.g., Sec.~8.3.1 of DM09) 
\begin{equation}
	f_{\gamma}(x)=\dfrac{1}{2\beta_{\pi}\gamma_{\pi} }H\left(x-\dfrac{1-\beta_{\pi}}{2}\right)H\left(\dfrac{1+\beta_{\pi}}{2}-x\right).
\end{equation}
The neutrino distribution function from the secondary muon decay is
\begin{equation}
	\begin{aligned}
		f_{\mbox{$\bar{\nu}$}_{\mbox{\scriptsize{$\mu$}}} }(x,h) &=
		f_{\mbox{$\nu$}_{\mbox{\scriptsize{$\mu$}}} }(x,-h) =
		\left(\dfrac{5}{3}-3x^{2}+\dfrac{4}{3}x^{3}\right)+h\times\left(-\dfrac{1}{3}+3x^{2}-\dfrac{8}{3}x^{3}\right)\\
		f_{\mbox{$\nu$}_{\mbox{\scriptsize{$e$}}} }(x,h) &=
		f_{\mbox{$\bar{\nu}$}_{\mbox{\scriptsize{$e$}}} }(x,-h) =
		(2-6x^{2}+4x^{3})+h\times(2-12x+18x^{2}-8x^{3})
	\end{aligned}
\end{equation}
and for $e^{\pm}$ (now omitting chiralities) the distribution can be simply expressed as \citep[cf., e.g.,][]{ParticleDataGroup:2024cfk}
\begin{equation}
	f_{e^\pm}(x)=\dfrac{4}{3}(1-x^{3})
\end{equation}
under the relativistic approximation, $\gamma_{e^\pm}\gtrsim10$. For the parent particle $X$ (that is always a massive particle with Lorentz factor $\gamma_X$) a sink term is implemented:

\begin{equation}
    \alpha_{X, \mathrm{dec}} (\gamma_X) = \frac{1}{\gamma_X \: t_{X,\mathrm{dec}}}.
\end{equation}

\subsection{Injection and escape}
\subsubsection*{Injection}
Although in principle arbitrary particle distributions can be injected directly through arrays, \AM offers built-in functions to inject a broken power-law electron and proton distribution with an exponential cutoff. 
Once the power-law properties such as spectral index and minimum/break/maximum energy are defined, the spectra are normalized to a (user-given) injected power per unit volume. \\
It is further possible to estimate the maximum electron/proton energy by balancing acceleration and losses. For this, an acceleration timescale as a fraction of the Bohm efficiency for the electrons/protons needs to be specified. The total cooling timescale in each energy grid is defined as the sum of all cooling processes included in the specific calculation.

\subsubsection*{Escape}
The following simple escape mechanisms are implemented: 

\begin{enumerate}
    \item No escape: All particles are confined and the escape term is thus equal to zero for all species.
    \item Energy-independent escape: For each particle species, the escape timescale can be defined as a fraction $\eta_X$ of the user-defined escape timescale $t_\mathrm{esc}$: 
    \begin{equation}
        \alpha_{X, \mathrm{esc}} (E_X) = \eta_X/t_\mathrm{esc}
    \end{equation}.
    \item Charged/uncharged particles: For convenience, a function to set the energy-independent escape timescale for all charged/uncharged particles is also included.
\end{enumerate}

\section{Implementation of the numerical solver}
\label{app:solver}
Below we provide further details on  the different solver regimes and their implementation. 
The threshold values for switching between the different regimes of the solver can be adjusted with three (user-adjustable) variables: the sink regime condition is 
$A/\alpha <$ \texttt{solver\_theshold\_esc\_dom}, the condition for the tri-diagonal matrix regime is $A \, \Delta t/\Delta x<$ \texttt{solver\_theshold\_matrix} and for the advection regime it is $A /\alpha <$ \texttt{solver\_theshold\_cool\_dom}.

\subsection{Sink regime ($A(x)\ll \alpha(x)$)}
In the sink regime, the advection term $A(x)$ is negligible compared to the sink term $\alpha(x)$ and \equ{overalllogarithmicform} can be approximated as
\begin{equation}
	\partial_{t}n(x,t)=-\alpha(x,t)n(x,t)+\epsilon(x,t) \, .
\end{equation}
To calculate the particle spectrum after a small time step $\Delta t$ we assume that the coefficients $\alpha(x)$ and $\epsilon(x)$ are constant in time and the analytical solution can be used:
\begin{equation}
    n(x, t+\Delta t) = n(x, t) e^{-\alpha(x,t) \Delta t} + \frac{\epsilon(x,t)}{\alpha(x, t)} \left( 1 - e^{-\alpha(x,t) \Delta t} \right).
\end{equation}
Note that species without advection terms like photons, neutrons and neutrinos always fall in this regime.

\subsection{Advection regime ($A(x) \gg \alpha(x)$) and sink-advection regime ($A(x) \approx \alpha(x), A(x) \gg \Delta x/\Delta t$)}
If the advection term is non-negligible and large compared to $\Delta x / \Delta t$ (Courant-Friedrichs-Lewy stability criterion), the treatment is similar to the sink regime. Assuming that the coefficients $A(x)$, $\alpha(x)$, $\epsilon(x)$ remain constant over a small time step $\Delta t$, we again use the analytical solution of \equ{overalllogarithmicform} 
\begin{equation}
	n(x,t+\Delta t)=\dfrac{A(y,t)}{A(x,t)} \exp \left[-\int_{y}^{x}\dfrac{\alpha(x^{\prime},t)}{A(x^{\prime},t)}dx^{\prime}\right]n(y,t)+\dfrac{1}{A(x,t)}\int_{y}^{x} dx^{\prime}\epsilon(x^{\prime},t)\exp\left[-\int_{x^{\prime}}^{x}\dfrac{\alpha(x^{\prime\prime},t)}{A(x^{\prime\prime},t)}dx^{\prime\prime}\right]
    \label{eq:semianalyticAalpha}
\end{equation}
where $y$ is the solution of the equation
\begin{equation}
    -\int_{x}^{y}\dfrac{dx^{\prime}}{A(x^{\prime},t)} = \Delta{t}\, 
\end{equation}
that can be calculated numerically for every energy $x$. 
We point out that this is the slowest solver, as evaluating the Green's function requires solving the integral $-\int_{y}^{x}\dfrac{\alpha(x^{\prime},t)}{A(x^{\prime},t)}dx^{\prime}$.
We therefore treat the case $\alpha(x) \ll A(x)$ (the advection regime) separately and Eq. \ref{eq:semianalyticAalpha} simplifies significantly for $\alpha(x) = 0$:
\begin{equation} \label{eq:semianalyticA}
	n(x,t+\Delta t)=\dfrac{A(y,t)}{A(x,t)}n(y,t)+\dfrac{1}{A(x,t)}\int_{y}^{x} dx^{\prime}\epsilon(x^{\prime},t) \; .
\end{equation}

\subsection{Tri-diagonal matrix regime ($A(x) \approx \alpha(x), A(x) \ll \Delta x/\Delta t$)}
Finally, if $A \ll \Delta x / \Delta t$ and sink and cooling terms are of the same order of magnitude, \AM makes use of the tri-diagonal matrix representation of the PDEs. This allows for efficient solving with the Thomas algorithm, where the computation time is significantly smaller than for the semi-analytic strategy.
We point out that the reasons for limiting the use of the matrix solver to the regime of $A \ll \Delta x / \Delta t$ (Courant-Friedrichs-Lewy stability criterion) are potential stability issues: outside this regime, particles advect by more than one energy bin within a time step, which cannot be treated accurately with a tri-diagonal solver.

Using the abbreviated notations from above (recall that the subscripts $i$ refer to the energy bin, while superscripts $k$ indicate the time step) we define for a quantity $s$

\begin{equation}
s\equiv s_{i}^{k},~s_{1}\equiv s_{i+1},~s_{-1}\equiv s_{i-1},~s^{1}\equiv s^{k+1}\, . 
\end{equation}
The discretization scheme is chosen as follows \citep[see, e.g.,][]{1970:ChangCooper,1996:ParkPetrosian}:

\begin{equation}
	\frac{n^{1}-n}{\Delta{t}}= - \frac{F_{1}^{+}-F_{-1}^{+}}{\Delta{x}} - \alpha (n^{1}-n)+\epsilon
\end{equation}
where we define the discretization of the flux $F(x) = A(x)$:

\begin{equation}
	F_{1}^{+} =A_{1}n_{1}^{+} \, ,  F_{-1}^{+} =A_{-1}n_{-1}^{+} \, .
\end{equation}
Note that in contrast to G16, the current implementation does not contain diffusion terms, which in the Chang \& Cooper formalism implies that the weight $\delta \rightarrow 0$. This leads to the above, simplified expressions for the fluxes. The time half-grid steps $n^+$ introduced above are chosen following a Crank-Nicolson scheme:
\begin{equation}
	n^{+}=(n^{1}+n)/2 \, .
\end{equation}

\end{CJK*}
\end{document}